\documentclass[apj]{emulateapj}

\usepackage{lscape}

\newcommand{\etal}{{et al.~}}
\newcommand{\msunh}{\>h^{-1}\rm M_\odot}
\newcommand{\Msun}{\>{\rm M_\odot}}

\newcommand{\Lsun}{\>{\rm L_{\odot}}}
\newcommand{\mpch}{\>h^{-1}{\rm {Mpc}}}

\newcommand{\calC}{{\cal C}}
\newcommand{\rmag}{\>^{0.1}{\rm M}_r-5\log h}
\newcommand{\rrmag}{\>^{0.0}{\rm M}_r-5\log h}

\shorttitle{Galaxy Groups in SDSS DR4: III}

\shortauthors{Yang et al.}

\begin{document}


\title{Galaxy Groups in the SDSS DR4: III. the luminosity and stellar mass
  functions}

\author{Xiaohu Yang\altaffilmark{1,4}, H.J. Mo \altaffilmark{2}, Frank C. van
  den Bosch\altaffilmark{3}}

\altaffiltext{1}{Shanghai Astronomical Observatory, the Partner Group of MPA,
  Nandan Road 80, Shanghai 200030, China; E-mail: xhyang@shao.ac.cn}

\altaffiltext{2}{Department of Astronomy, University of Massachusetts, Amherst
  MA 01003-9305}

\altaffiltext{3} {Max-Planck-Institute for Astronomy, K\"onigstuhl 17, D-69117
  Heidelberg, Germany }

\altaffiltext{4}{Joint Institute for Galaxy and Cosmology (JOINGC) of Shanghai
  Astronomical Observatory and University of Science and Technology of China}


\begin{abstract}  Using a large  galaxy group  catalogue constructed  from the
  Sloan  Digital  Sky  Survey Data  Release  4  (SDSS  DR4) with  an  adaptive
  halo-based  group finder,  we investigate  the luminosity  and  stellar mass
  functions for  different populations of galaxies  (central versus satellite;
  red versus blue; and galaxies in  groups of different masses) and for groups
  themselves. The  conditional stellar  mass function (CSMF),  which describes
  the stellar  distribution of galaxies in  halos of a given  mass for central
  and satellite galaxies  can be well modeled with  a log-normal distribution
  and a modified  Schechter form, respectively. On average,  there are about 3
  times  as  many  central   galaxies  as  satellites.   Among  the  satellite
  population, there are  in general more red galaxies than  blue ones. For the
  central population, the luminosity function  is dominated by red galaxies at
  the massive  end, and by  blue galaxies  at the low  mass end.  At  the very
  low-mass  end ($M_\ast  \la 10^9  h^{-2}\Msun$), however, there is a  marked
  increase in  the number of red  centrals.  We speculate  that these galaxies
  are located close  to large halos so that their  star formation is truncated
  by the large-scale environments.  The stellar-mass function of galaxy groups
  is well described by a double  power law, with a characteristic stellar mass
  at $\sim 4\times 10^{10}h^{-2}\Msun$.   Finally, we use the observed stellar
  mass function of central galaxies to  constrain the stellar mass - halo mass
  relation for low  mass halos, and obtain $M_{\ast,  c}\propto M_h^{4.9}$ for
  $M_h \ll 10^{11} \msunh$.
\end{abstract}




\section{Introduction}

In recent years,  great progress has been made in  our understanding about how
galaxies form and evolve in dark matter halos owing to the development of halo
models  and   the  related  halo  occupation  models.    The  halo  occupation
distribution (hereafter HOD), $P(N \vert M_h)$, which gives the probability of
finding $N$ galaxies (with some specified properties) in a halo of mass $M_h$,
has  been extensively used  to study  the galaxy  distribution in  dark matter
halos and galaxy clustering on large  scales (e.g.  Jing, Mo \& B\"orner 1998;
Peacock \& Smith 2000; Seljak  2000; Scoccimarro \etal 2001; Jing, B\"orner \&
Suto 2002;  Berlind \&  Weinberg 2002; Bullock,  Wechsler \&  Somerville 2002;
Scranton 2002; Zehavi  \etal 2004, 2005; Zheng \etal  2005; Tinker \etal 2005;
Skibba et al. 2007; Brown  et al.  2008).  The conditional luminosity function
(hereafter CLF), $\Phi(L \vert M_h) {\rm d}L$, which refines the HOD statistic
by  considering the average  number of  galaxies with  luminosity $L  \pm {\rm
d}L/2$  that reside  in  a halo  of  mass $M_h$,  has  also been  investigated
extensively (Yang, Mo \&  van den Bosch 2003; van den Bosch,  Yang \& Mo 2003;
Vale \& Ostriker 2004, 2008; Cooray 2006;  van den Bosch et al.  2007) and has
been applied  to various redshift surveys,  such as the  2-degree Field Galaxy
Redshift Survey (2dFGRS), the Sloan  Digital Sky Survey (SDSS) and DEEP2 (e.g.
Yan, Madgwick \& White 2003; Yang \etal 2004; Mo et al. 2004; Wang \etal 2004;
Zehavi \etal 2005; Yan, White \& Coil 2004).  These investigations demonstrate
that the HOD/CLF statistics are  very powerful tools to establish and describe
the  connection between galaxies  and dark  matter halos,  providing important
constraints  on  various physical  processes  that  govern  the formation  and
evolution of  galaxies, such as  gravitational instability, gas  cooling, star
formation, merging,  tidal stripping  and heating, and  a variety  of feedback
processes, and how their efficiencies scale with halo mass.  Furthermore, they
also  indicate that  the  galaxy/dark halo  connection  can provide  important
constraints  on cosmology  (e.g.,van  den Bosch,  Mo  \& Yang  2003; Zheng  \&
Weinberg 2007).

However, as pointed out in Yang et al.  (2005c; hereafter Y05c), a shortcoming
of  the  HOD/CLF  models  is   that  the  results  are  not  completely  model
independent.  Typically, assumptions have  to be made regarding the functional
form of  either $P(N  \vert M_h)$  or $\Phi(L \vert  M_h)$.  Moreover,  in all
HOD/CLF studies to date, the  occupation distributions have been determined in
an  indirect way:  the  free parameters  of  the assumed  functional form  are
constrained  using {\it  statistical}  data on  the  abundance and  clustering
properties of the galaxy population.   An alternative method that can directly
probe the galaxy - dark halo connection (e.g. HOD/CLF models) is to use galaxy
groups as  a representation of dark matter  halos and to study  how the galaxy
population changes with  the properties of the groups  (e.g., Y05c; Zandivarez
et al. 2006; Robotham  et al. 2006; Hansen \etal 2007; Yang  et al. 2008). For
such a  purpose, one has to properly  find the galaxy groups  that are closely
connected to  the dark matter halos.   In recent studies, Yang  et al. (2005a;
2007) developed  an adaptive  halo-based group finder  that has  such features
\footnote{In this paper, we refer  to systems of galaxies as groups regardless
  of their richness, including isolated  galaxies (i.e., systems with a single
  member) and rich clusters of galaxies.}.  This group finder has been applied
to the  2dFGRS (Yang et al.   2005a) and to  the SDSS (Weinmann et  al. 2006a;
Yang et al. 2007).  Detailed tests with mock galaxy catalogues have shown that
this  group finder  is very  successful in  associating galaxies  according to
their  common dark  matter halos.   In particular,  the group  finder performs
reliably  not only  for rich  systems, but  also for  poor  systems, including
isolated central galaxies in low mass  halos.  This makes it possible to study
the  galaxy-halo connection  for systems  covering  a large  dynamic range  in
masses.  With  a well-defined  galaxy group catalogue,  one can then  not only
study the properties  of galaxies in different groups  (e.g.  Y05c; Yang \etal
2005d; Collister \& Lahav 2005; van den Bosch \etal 2005; Robotham \etal 2006;
Zandivarez  \etal 2006;  Weinmann \etal  2006a,b;  van den  Bosch \etal  2008;
McIntosh \etal 2007; Yang et al.   2008), but also probe how dark matter halos
trace the large-scale structure of the Universe (e.g.  Yang \etal 2005b, 2006;
Coil \etal 2006; Berlind \etal 2007; Wang et al.  2008a).

Recently, this group finder has been applied to the Sloan Digital Sky Survey
Data Release 4 (SDSS DR4), and the group catalogues constructed are described
in detail in Yang \etal (2007; Paper I hereafter).  In these catalogues
various observational selection effects are taken into account, and each of
the groups is assigned a reliable halo mass. The group catalogues including
the membership of the groups are available at these links
\footnote{http://gax.shao.ac.cn/data/Group.html}
\footnote{http://www.astro.umass.edu/$^\sim$xhyang/Group.html}.  In Yang et
al.  (2008; Paper II hereafter) we have used these group catalogues to obtain
various halo occupation statistics and to measure the CLFs for different
populations of galaxies. In this paper, the third in the series, we will focus
on the conditional stellar mass functions (CSMFs) for different populations of
galaxies. In addition, we will also examine the general luminosity and stellar
mass functions for different populations of galaxies and for groups
themselves.  Finally, we will demonstrate how to use the observed luminosity
and stellar mass functions for central galaxies to constrain the HOD in small
halos.

This paper is organized as  follows: In Section~\ref{sec_data} we describe the
data (galaxy and group catalogues) used in this paper. Section~\ref{sec_CSMFs}
presents  our  measurement of  the  CSMFs  for all,  red  and  blue galaxies.  
Sections~\ref{sec_LF_gax} and ~\ref{sec_LF_grp} present our measurement of the
luminosity and stellar  mass functions for galaxies and  groups, respectively. 
In Section  \ref{sec_small}, we probe  the properties of the  central galaxies
that can be formed in those small halos.  Finally, we summarize our results in
Section~\ref{sec_summary}.   Throughout  this  paper,  we use  a  $\Lambda$CDM
`concordance' cosmology  whose parameters  are consistent with  the three-year
data    release    of    the     WMAP    mission:    $\Omega_m    =    0.238$,
$\Omega_{\Lambda}=0.762$,  $n_s=0.951$, $h=0.73$ and  $\sigma_8=0.75$ (Spergel
et al. 2007).  If not quoted, the units of luminosity, stellar and halo masses
are in terms of  $h^{-2}\Lsun$, $h^{-2}\Msun$ and $h^{-1}\Msun$, respectively. 
Finally, unless  noted differently, the luminosity functions  and stellar mass
functions are  presented in units  of $h^3{\rm Mpc}^{-3}  {\rm d} \log  L$ and
$h^3{\rm Mpc}^{-3} {\rm  d} \log M_{\ast}$, respectively, where  $\log$ is the
10 based logarithm.

\section{Data}
\label{sec_data}

\subsection{Galaxy and group catalogues}

The data  used in our analysis  here are the same  as those used  in Paper II.
Readers who  have already read  through Paper II  may go directly to  the next
subsection.

The group catalogues are constructed  from the New York University Value-Added
Galaxy Catalogue (NYU-VAGC;  Blanton \etal 2005b), which is  based on the SDSS
Data Release 4  (Adelman-McCarthy \etal 2006), but with  an independent set of
significantly improved  reductions.  From NYU-VAGC  we select all  galaxies in
the Main Galaxy Sample with redshifts in the range $0.01 \leq z \leq 0.20$ and
with a  redshift completeness $\calC >  0.7$.  As described in  Paper I, three
group samples are constructed from the corresponding galaxy samples: Sample I,
which only  uses the $362356$  galaxies with measured $r$-band  magnitudes and
redshifts from the SDSS, Sample II which also includes 7091 galaxies with SDSS
$r$-band magnitudes  but redshifts taken from alternative  surveys, and Sample
III which includes  an additional $38672$ galaxies that  lack redshifts due to
fiber  collisions  but  that  are  assigned the  redshifts  of  their  nearest
neighbors. Although this  fiber collision correction works well  in roughly 60
percent of all cases, the remaining 40 percent are assigned redshifts that can
be very different from their true  values (Zehavi \etal 2002).  Samples II and
III should  therefore be considered as two  extremes as far as  a treatment of
fiber-collisions is concerned. Unless  stated otherwise, our results are based
on  Sample II.  For  comparison, we  also present  some results  obtained from
Sample III.

The  magnitudes and  colors of  all galaxies  are based  on the  standard SDSS
Petrosian technique (Petrosian 1976;  Strauss \etal 2002), have been corrected
for galactic  extinction (Schlegel, Finkbeiner  \& Davis 1998), and  have been
$K$-corrected and  evolution corrected to $z=0.1$, using  the method described
in  Blanton \etal  (2003a; b).  We use  the notation  $\rmag$ to  indicate the
resulting absolute magnitude in the  $r$-band. The galaxies are separated into
red  and blue  subsamples according  to  their bi-normal  distribution in  the
$^{0.1}(g-r)$ color (Baldry  \etal 2004; Blanton \etal 2005a;  Li \etal 2006),
using the separation criteria (see Paper II),
\begin{equation}\label{quadfit}
^{0.1}(g-r) = 1.022-0.0651x-0.00311x^2\,,
\end{equation}
where $x=\rmag + 23.0$.

Stellar masses,  indicated by $M_*$, for  all galaxies are  computed using the
relations  between stellar  mass-to-light ratio  and $^{0.0}(g-r)$  color from
Bell \etal (2003),
\begin{eqnarray} \label{eq:stellar}
\log\left[{M_* \over h^{-2}\Msun}\right] & = & -0.306 + 1.097
\left[^{0.0}(g-r)\right] - 0.10 \nonumber\\
& & - 0.4(\rrmag-4.64)\,.
\end{eqnarray}
Here $^{0.0}(g-r)$ and  $\rrmag$ are the $(g-r)$ color  and $r$-band magnitude
$K+E$ corrected to $z=0.0$; $4.64$ is the $r$-band magnitude of the Sun in the
AB system (Blanton  \& Roweis 2007); and the  $-0.10$ term effectively implies
that we adopt a Kroupa (2001) IMF (Borch \etal 2006).

For each group in our catalogue we  have two estimates of its dark matter halo
mass $M_h$:  (1) $M_L$, which  is based on  the ranking of  the characteristic
group luminosity $L_{19.5}$ , and (2)  $M_S$, which is based on the ranking of
the     characteristic    group     stellar     mass    $M_{\rm     stellar}$,
respectively\footnote{$L_{19.5}$ and $M_{\rm  stellar}$ are, respectively, the
total luminosity and total stellar mass  of all group members with $\rmag \leq
-19.5$.}. The halo  mass is estimated for each group with  at least one member
galaxy that is brighter than $\rmag =  -19.5$.  As shown in Paper I, these two
halo masses agree reasonably well with each other, with scatter that decreases
from $\sim 0.1$ dex at the low-mass end to $\sim 0.05$ dex at the massive end.
Detailed tests using  mock galaxy redshift surveys have  demonstrated that the
group masses thus estimated can recover the true halo masses with a 1-$\sigma$
deviation of  $\sim 0.3$ dex,  and are more  reliable than those based  on the
velocity dispersion of group members (Y05c; Weinmann \etal 2006; Berlind \etal
2006;  Paper I).   Note also  that survey  edge effects  have been  taken into
account in our group catalogue:  groups that suffer severely from edge effects
(about 1.6\%  of the  total) have  been removed from  the catalogue.   In most
cases, we take the  most massive galaxy (in terms of stellar  mass) in a group
as  the  central  galaxy (MCG)  and  all  others  as satellite  galaxies.   In
addition, we also considered a case in which the brightest galaxy in the group
is considered as the central galaxy  (BCG).  Tests have shown that for most of
what follows, these two definitions yield indistinguishable results.  Whenever
the two  definitions lead to  significant differences, we present  results for
both. Throughout  this paper, results are  calculated for both  samples II and
III using  both halo masses, $M_L$  and $M_S$. Any  significant differences in
the  results due  to  the use  of  different samples  and  mass estimates  are
discussed.

Finally,  we  caution that  the  SDSS  pipeline  may have  underestimated  the
luminosities for  bright galaxies  (e.g. von  der Linden et  al. 2007;  Guo et
al.  2009). According  to  Guo et  al.  the  NYU-VAGC magnitude  is
overestimated by  about $0.5\pm 0.1$  at apparent magnitudes $r\sim  13.0$ and
about $0.1\pm  0.1$ at $r\sim  17.0$. Although this  will not change  the halo
masses  estimated  using  the  abundance  match to  halo  mass  function,  the
luminosity  and stellar  mass functions  for galaxies  and groups  are shifted
slightly at the bright ends if a correction is made to the SDSS luminosities.

\subsection{Galaxy and group completeness limits}\label{sec:comp}

\begin{figure} \plotone{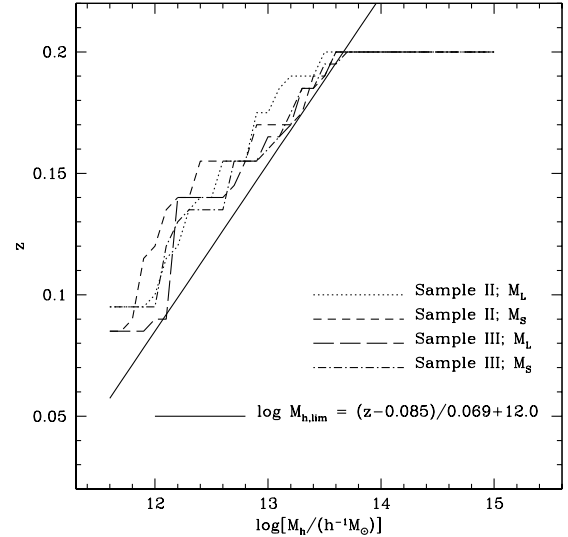}
  \caption{The group completeness limit as a function of halo mass.  Different
    lines  (except the  solid  one) correspond  to  different combinations  of
    Samples  and halo  masses estimated  ($M_L$ or  $M_S$) as  indicated.  The
    solid  line illustrates  a  conservative  halo mass  limit  for all  those
    combinations.  That  is the  halos below given  redshift with  halo masses
    larger than $M_{h,{\rm lim}}$ are complete.  }
\label{fig:Mh_limit}
\end{figure}

Because of the  survey magnitude limit, only bright galaxies  can be observed. 
This consequently  will induce incompleteness in the  distribution of galaxies
with respect to  absolute magnitude and stellar mass,  and in the distribution
of groups with halo mass. In this subsection, we discuss such completeness and
how to make corrections.

As discussed  in detail in the  Appendix of van  den Bosch et al.  (2008), the
apparent magnitude limit of the  galaxy sample ($m_r=17.77$) can be translated
into a redshift-dependent absolute magnitude limit given by
\begin{eqnarray} \label{magn01r}
\lefteqn{ ^{0.1}{\rm M}_{r,{\rm lim}} - 5\log h = } \nonumber \\
& & 17.77 - {\rm DM}(z) - k_{0.1}(z) + 1.62(z-0.1)\,.
\end{eqnarray}
where $k_{0.1}(z)$ is the $K$-correction to $z=0.1$, the $1.62(z-0.1)$ term is
the evolution correction of Blanton \etal (2003b), and
\begin{equation} \label{distmeas}
{\rm DM}(z) = 5 \log D_L(z) + 25
\end{equation}
is  the distance module corresponding  to  redshift $z$,  with $D_L(z)$  the
luminosity distance  in $\mpch$.  Using  the $K$-corrections of  Blanton \etal
(2003a;  see  also  Blanton  \&  Roweis  2007),  the  redshift-dependence  is
reasonably well described by
\begin{equation} \label{kcorrect}
k_{0.1}(z) = 2.5\log\left({z+0.9 \over 1.1}\right)\,.
\end{equation}
At $z=0.1$,  where (by definition)  $k_{0.1} = -2.5\log(1.1) \simeq  -0.1$ for
all galaxies (e.g.,  Blanton \& Roweis 2007), this  exactly gives the absolute
magnitude  limit of the  sample.  At  lower and  higher redshifts,  however, a
small fraction of the sample galaxies fall below this limit.  This owes to the
fact that  $k_{0.1}(z)$ not only  depends on redshift  but also on  color.  In
order to ensure completeness, we use a conservative absolute-magnitude limit:
\begin{equation} \label{eq:maglim}
{^{0.1}{\rm M}'_{r,{\rm lim}} - 5\log h} = 
{^{0.1}{\rm M}_{r,{\rm lim}} - 5\log h} - 0.1\,,
\end{equation} 
where the term $0.1$ takes into account the scatter in the $K$ correction. For
the stellar masses of the SDSS galaxies, we adopt, for given redshift $z$, the
following completeness limit:
\begin{eqnarray} \label{eq:mstarlim}
\lefteqn{\log[M_{*,{\rm lim}}/(h^{-2}\Msun)] =} \\
 & & {4.852 + 2.246 \log D_L(z) + 1.123 \log(1+z) - 1.186 z \over 1 - 0.067
  z} \,. \nonumber
\end{eqnarray}
(see van den Bosch et al. 2008)

Next we consider incompleteness in the actual group catalogue.  As illustrated
in  Fig. 6 of  Yang et  al.  (2007),  the groups  within a  certain luminosity
$L_{19.5}$  or stellar  mass $M_{\rm  stellar}$  bin (which  corresponds to  a
certain halo mass  bin) are complete only to a  certain redshift.  Beyond this
redshift the  number density  of the groups  drops dramatically.  We therefore
need to  take this completeness  into account. Here  we proceed as  follows to
obtain the halo-mass  completeness limit at any given  redshift.  First, for a
given halo  mass $M_h$, we  measure the number  densities of groups  with halo
mass  within   $\log  M_h\pm   0.05$,  $n(0.01,z_{\rm  max})$   and  $n(z_{\rm
  max}-\Delta z,z_{\rm  max})$, within  the redshift ranges,  [$0.01$, $z_{\rm
  max}$] and  [$z_{\rm max}-\Delta z$, $z_{\rm max}$],  respectively.  Here we
set $\Delta z=0.005$. If the group  sample is not complete at redshift $z_{\rm
  max}$, then  the group number density $n(z_{\rm  max}-\Delta z,z_{\rm max})$
is  expected to  drop significantly.   Starting  from $z_{\rm  max} =0.2$,  we
iteratively   decrease  $z_{\rm   max}$  according   to   $z_{\rm  max}=z_{\rm
  max}-\Delta z$, and find the largest redshift $z_{\rm max}$ that satisfies,
\begin{equation}\label{eq:n_z}
n(z_{\rm max}-\Delta z,z_{\rm max})+ 3\sigma_n \ge n(0.01,z_{\rm  max}) \,,
\end{equation}
where  $\sigma_n$ is  the  variance in  the  number density  of groups  within
[$z_{\rm  max}-\Delta z$,  $z_{\rm  max}$] among  200  bootstrap samples.  The
$z_{\rm max}$ thus obtained is the one at which the halos are complete down to
the corresponding halo mass $M_h$.  For  the group samples used in this paper,
Samples II  and III, and  for the  two sets of  halo masses estimated  for our
groups, $M_L$ and $M_S$, we obtain the value of $z_{\rm max}$ as a function of
halo mass. The results are  shown in Fig.~\ref{fig:Mh_limit}.  From this plot,
we obtain a conservative halo-mass limit,
\begin{equation}\label{eq:Mh_limit}
\log M_{h, {\rm lim}} =(z-0.085)/0.069 + 12\,,
\end{equation}
which is  shown as the  solid line in Fig.~\ref{fig:Mh_limit}.   Clearly, this
criterion works  for both Samples II  and III, as  well as for both  $M_L$ and
$M_S$. Thus,  for a  given redshift  $z$, groups with  masses $\ge  M_{h, {\rm
lim}}$ are complete.

\section{The galaxy luminosity and stellar mass functions: central vs
  satellite}
\label{sec_LF_gax}

\begin{turnpage}
\begin{deluxetable*}{cccccccccccc} 
  \tabletypesize{\scriptsize} 
  \tablecaption{The galaxy luminosity functions $\Phi(L)$}
  \tablewidth{0pt} 
  \tablehead{  &\multicolumn{3}{c}{ALL} && \multicolumn{3}{c}{CENTRAL} && \multicolumn{3}{c}{SATELLITE}\\ 
  \cline{2-4} \cline{6-8} \cline{10-12}\\ (1) & (2) & (3) & (4) & & (5) & (6) &
  (7) && (8) & (9) & (10) \\
   $\log L$ & all & red & blue && all & red & blue && all & red & blue } 

\startdata
  7.8  &  9.0324 $\pm$ 3.3897  &  5.3944 $\pm$ 2.6814  &  3.6379 $\pm$ 2.0358  &&  7.0216 $\pm$ 3.2849  &  4.1403 $\pm$ 2.6068  &  2.8813 $\pm$ 1.9104  &&  2.0108 $\pm$ 1.6054  &  1.2542 $\pm$ 0.9849  &  0.7566 $\pm$ 1.0084 \\
  7.9  &  7.5462 $\pm$ 1.3850  &  4.5484 $\pm$ 1.1735  &  2.9977 $\pm$ 0.7789  &&  4.8617 $\pm$ 1.5245  &  2.4132 $\pm$ 1.0153  &  2.4485 $\pm$ 0.8946  &&  2.6845 $\pm$ 1.5813  &  2.1353 $\pm$ 1.3264  &  0.5492 $\pm$ 0.5109 \\
  8.0  &  6.3651 $\pm$ 0.8115  &  2.8845 $\pm$ 0.5102  &  3.4806 $\pm$ 0.6290  &&  4.5706 $\pm$ 0.8858  &  2.0541 $\pm$ 0.5558  &  2.5166 $\pm$ 0.6278  &&  1.7945 $\pm$ 0.8846  &  0.8305 $\pm$ 0.5089  &  0.9640 $\pm$ 0.5210 \\
  8.1  &  6.0813 $\pm$ 0.6498  &  2.5566 $\pm$ 0.3424  &  3.5247 $\pm$ 0.5074  &&  4.4285 $\pm$ 0.7806  &  1.7563 $\pm$ 0.3504  &  2.6723 $\pm$ 0.5992  &&  1.6528 $\pm$ 0.7226  &  0.8003 $\pm$ 0.2809  &  0.8524 $\pm$ 0.5526 \\
  8.2  &  4.2781 $\pm$ 0.4229  &  1.8413 $\pm$ 0.2858  &  2.4368 $\pm$ 0.2576  &&  2.7779 $\pm$ 0.5448  &  1.0132 $\pm$ 0.2620  &  1.7647 $\pm$ 0.3909  &&  1.5002 $\pm$ 0.5502  &  0.8281 $\pm$ 0.3306  &  0.6721 $\pm$ 0.3188 \\
  8.3  &  4.9823 $\pm$ 0.4415  &  1.8511 $\pm$ 0.2813  &  3.1312 $\pm$ 0.2737  &&  3.4498 $\pm$ 0.5347  &  1.0388 $\pm$ 0.1995  &  2.4110 $\pm$ 0.4263  &&  1.5326 $\pm$ 0.5281  &  0.8124 $\pm$ 0.2851  &  0.7202 $\pm$ 0.3201 \\
  8.4  &  4.5373 $\pm$ 0.3176  &  1.4754 $\pm$ 0.1732  &  3.0620 $\pm$ 0.2409  &&  3.0877 $\pm$ 0.4414  &  0.7873 $\pm$ 0.1432  &  2.3004 $\pm$ 0.3883  &&  1.4496 $\pm$ 0.4447  &  0.6881 $\pm$ 0.1755  &  0.7615 $\pm$ 0.3339 \\
  8.5  &  4.7646 $\pm$ 0.2712  &  1.4727 $\pm$ 0.1820  &  3.2919 $\pm$ 0.1862  &&  3.2717 $\pm$ 0.3598  &  0.8272 $\pm$ 0.0833  &  2.4445 $\pm$ 0.3416  &&  1.4930 $\pm$ 0.3578  &  0.6455 $\pm$ 0.1481  &  0.8474 $\pm$ 0.2594 \\
  8.6  &  5.0611 $\pm$ 0.1803  &  1.3903 $\pm$ 0.1616  &  3.6708 $\pm$ 0.2012  &&  3.4916 $\pm$ 0.3415  &  0.6852 $\pm$ 0.0744  &  2.8064 $\pm$ 0.3192  &&  1.5695 $\pm$ 0.3034  &  0.7051 $\pm$ 0.1422  &  0.8644 $\pm$ 0.2057 \\
  8.7  &  4.9694 $\pm$ 0.1665  &  1.5349 $\pm$ 0.1161  &  3.4344 $\pm$ 0.1741  &&  3.1904 $\pm$ 0.3180  &  0.6093 $\pm$ 0.0819  &  2.5811 $\pm$ 0.2763  &&  1.7790 $\pm$ 0.2697  &  0.9256 $\pm$ 0.1269  &  0.8534 $\pm$ 0.1816 \\
  8.8  &  4.7649 $\pm$ 0.1504  &  1.3059 $\pm$ 0.1176  &  3.4589 $\pm$ 0.1882  &&  3.0336 $\pm$ 0.3153  &  0.5235 $\pm$ 0.0731  &  2.5102 $\pm$ 0.2791  &&  1.7313 $\pm$ 0.2625  &  0.7825 $\pm$ 0.1343  &  0.9488 $\pm$ 0.1568 \\
  8.9  &  4.2500 $\pm$ 0.1426  &  1.2642 $\pm$ 0.0785  &  2.9857 $\pm$ 0.1575  &&  2.7792 $\pm$ 0.2663  &  0.5656 $\pm$ 0.0578  &  2.2136 $\pm$ 0.2340  &&  1.4708 $\pm$ 0.2005  &  0.6987 $\pm$ 0.0876  &  0.7721 $\pm$ 0.1348 \\
  9.0  &  3.7151 $\pm$ 0.1449  &  1.0853 $\pm$ 0.0615  &  2.6298 $\pm$ 0.1556  &&  2.4177 $\pm$ 0.2373  &  0.4832 $\pm$ 0.0559  &  1.9345 $\pm$ 0.1985  &&  1.2974 $\pm$ 0.1458  &  0.6021 $\pm$ 0.0827  &  0.6953 $\pm$ 0.0795 \\
  9.1  &  3.4459 $\pm$ 0.1257  &  1.0891 $\pm$ 0.0444  &  2.3568 $\pm$ 0.1293  &&  2.2779 $\pm$ 0.2120  &  0.5214 $\pm$ 0.0585  &  1.7565 $\pm$ 0.1743  &&  1.1680 $\pm$ 0.1280  &  0.5677 $\pm$ 0.0704  &  0.6003 $\pm$ 0.0713 \\
  9.2  &  3.2127 $\pm$ 0.1361  &  1.0609 $\pm$ 0.0291  &  2.1518 $\pm$ 0.1267  &&  2.1378 $\pm$ 0.1763  &  0.5271 $\pm$ 0.0433  &  1.6106 $\pm$ 0.1480  &&  1.0749 $\pm$ 0.0778  &  0.5337 $\pm$ 0.0439  &  0.5412 $\pm$ 0.0465 \\
  9.3  &  2.8792 $\pm$ 0.0998  &  1.0113 $\pm$ 0.0308  &  1.8679 $\pm$ 0.0993  &&  1.9424 $\pm$ 0.1626  &  0.5276 $\pm$ 0.0440  &  1.4148 $\pm$ 0.1316  &&  0.9367 $\pm$ 0.0924  &  0.4836 $\pm$ 0.0505  &  0.4531 $\pm$ 0.0507 \\
  9.4  &  2.7471 $\pm$ 0.0998  &  1.0501 $\pm$ 0.0278  &  1.6970 $\pm$ 0.0879  &&  1.8702 $\pm$ 0.1530  &  0.5831 $\pm$ 0.0500  &  1.2870 $\pm$ 0.1125  &&  0.8770 $\pm$ 0.0756  &  0.4670 $\pm$ 0.0434  &  0.4100 $\pm$ 0.0393 \\
  9.5  &  2.6091 $\pm$ 0.0972  &  1.0723 $\pm$ 0.0296  &  1.5368 $\pm$ 0.0783  &&  1.8150 $\pm$ 0.1525  &  0.6253 $\pm$ 0.0520  &  1.1897 $\pm$ 0.1094  &&  0.7941 $\pm$ 0.0729  &  0.4470 $\pm$ 0.0362  &  0.3471 $\pm$ 0.0430 \\
  9.6  &  2.6861 $\pm$ 0.0956  &  1.2200 $\pm$ 0.0332  &  1.4660 $\pm$ 0.0716  &&  1.8920 $\pm$ 0.1553  &  0.7424 $\pm$ 0.0642  &  1.1496 $\pm$ 0.0988  &&  0.7940 $\pm$ 0.0747  &  0.4776 $\pm$ 0.0433  &  0.3164 $\pm$ 0.0368 \\
  9.7  &  2.4858 $\pm$ 0.0875  &  1.2074 $\pm$ 0.0339  &  1.2784 $\pm$ 0.0616  &&  1.7868 $\pm$ 0.1444  &  0.7688 $\pm$ 0.0662  &  1.0180 $\pm$ 0.0856  &&  0.6990 $\pm$ 0.0697  &  0.4386 $\pm$ 0.0419  &  0.2604 $\pm$ 0.0315 \\
  9.8  &  2.1931 $\pm$ 0.0757  &  1.1136 $\pm$ 0.0334  &  1.0795 $\pm$ 0.0491  &&  1.6238 $\pm$ 0.1239  &  0.7524 $\pm$ 0.0579  &  0.8713 $\pm$ 0.0715  &&  0.5693 $\pm$ 0.0564  &  0.3611 $\pm$ 0.0320  &  0.2082 $\pm$ 0.0279 \\
  9.9  &  1.8351 $\pm$ 0.0654  &  0.9485 $\pm$ 0.0274  &  0.8866 $\pm$ 0.0428  &&  1.4036 $\pm$ 0.1038  &  0.6725 $\pm$ 0.0489  &  0.7311 $\pm$ 0.0593  &&  0.4315 $\pm$ 0.0450  &  0.2760 $\pm$ 0.0269  &  0.1555 $\pm$ 0.0205 \\
 10.0  &  1.5384 $\pm$ 0.0557  &  0.8246 $\pm$ 0.0255  &  0.7138 $\pm$ 0.0340  &&  1.1979 $\pm$ 0.0884  &  0.6080 $\pm$ 0.0452  &  0.5899 $\pm$ 0.0465  &&  0.3405 $\pm$ 0.0383  &  0.2165 $\pm$ 0.0241  &  0.1239 $\pm$ 0.0160 \\
 10.1  &  1.2460 $\pm$ 0.0444  &  0.6707 $\pm$ 0.0210  &  0.5753 $\pm$ 0.0265  &&  0.9992 $\pm$ 0.0702  &  0.5149 $\pm$ 0.0364  &  0.4843 $\pm$ 0.0366  &&  0.2468 $\pm$ 0.0297  &  0.1558 $\pm$ 0.0186  &  0.0910 $\pm$ 0.0124 \\
 10.2  &  0.9413 $\pm$ 0.0320  &  0.5112 $\pm$ 0.0160  &  0.4301 $\pm$ 0.0182  &&  0.7774 $\pm$ 0.0525  &  0.4108 $\pm$ 0.0274  &  0.3666 $\pm$ 0.0272  &&  0.1639 $\pm$ 0.0229  &  0.1004 $\pm$ 0.0133  &  0.0635 $\pm$ 0.0105 \\
 10.3  &  0.6428 $\pm$ 0.0226  &  0.3591 $\pm$ 0.0123  &  0.2837 $\pm$ 0.0120  &&  0.5448 $\pm$ 0.0353  &  0.3000 $\pm$ 0.0190  &  0.2448 $\pm$ 0.0177  &&  0.0980 $\pm$ 0.0141  &  0.0591 $\pm$ 0.0080  &  0.0389 $\pm$ 0.0067 \\
 10.4  &  0.3961 $\pm$ 0.0137  &  0.2302 $\pm$ 0.0084  &  0.1659 $\pm$ 0.0065  &&  0.3442 $\pm$ 0.0209  &  0.1995 $\pm$ 0.0121  &  0.1447 $\pm$ 0.0098  &&  0.0520 $\pm$ 0.0081  &  0.0307 $\pm$ 0.0046  &  0.0212 $\pm$ 0.0039 \\
 10.5  &  0.2227 $\pm$ 0.0075  &  0.1393 $\pm$ 0.0053  &  0.0834 $\pm$ 0.0030  &&  0.1981 $\pm$ 0.0104  &  0.1241 $\pm$ 0.0067  &  0.0740 $\pm$ 0.0044  &&  0.0246 $\pm$ 0.0035  &  0.0153 $\pm$ 0.0020  &  0.0094 $\pm$ 0.0018 \\
 10.6  &  0.1078 $\pm$ 0.0028  &  0.0733 $\pm$ 0.0030  &  0.0345 $\pm$ 0.0010  &&  0.0987 $\pm$ 0.0042  &  0.0674 $\pm$ 0.0036  &  0.0313 $\pm$ 0.0011  &&  0.0091 $\pm$ 0.0018  &  0.0059 $\pm$ 0.0009  &  0.0032 $\pm$ 0.0011 \\
 10.7  &  0.0475 $\pm$ 0.0010  &  0.0345 $\pm$ 0.0012  &  0.0130 $\pm$ 0.0011  &&  0.0447 $\pm$ 0.0012  &  0.0328 $\pm$ 0.0013  &  0.0119 $\pm$ 0.0006  &&  0.0027 $\pm$ 0.0008  &  0.0017 $\pm$ 0.0003  &  0.0011 $\pm$ 0.0006 \\
 10.8  &  0.0183 $\pm$ 0.0011  &  0.0140 $\pm$ 0.0006  &  0.0043 $\pm$ 0.0010  &&  0.0175 $\pm$ 0.0010  &  0.0135 $\pm$ 0.0006  &  0.0040 $\pm$ 0.0009  &&  0.0008 $\pm$ 0.0003  &  0.0005 $\pm$ 0.0001  &  0.0003 $\pm$ 0.0002 \\
 10.9  &  0.0060 $\pm$ 0.0006  &  0.0047 $\pm$ 0.0003  &  0.0013 $\pm$ 0.0004  &&  0.0059 $\pm$ 0.0005  &  0.0046 $\pm$ 0.0003  &  0.0013 $\pm$ 0.0004  &&  0.0001 $\pm$ 0.0001  &  0.0000 $\pm$ 0.0000  &  0.0000 $\pm$ 0.0001 \\
 11.0  &  0.0013 $\pm$ 0.0002  &  0.0011 $\pm$ 0.0002  &  0.0002 $\pm$ 0.0001  &&  0.0013 $\pm$ 0.0002  &  0.0011 $\pm$ 0.0002  &  0.0002 $\pm$ 0.0001  &&  0.0000 $\pm$ 0.0000  &  0.0000 $\pm$ 0.0000  &  0.0000 $\pm$ 0.0000 
\enddata
 
\tablecomments{Column  (1):  the  median   of  the  logarithm  of  the  galaxy
  luminosity with  bin width  $\Delta \log L  = 0.05$.   Columns (2 -  4): the
  luminosity functions of all, red  and blue galaxies for 'ALL' group members.
  Columns (5 - 7): the luminosity functions for all, red and blue galaxies for
  'CENTRAL' group members.  Columns (8 - 10): the luminosity functions of all,
  red and blue galaxies for 'SATELLITE' group members, respectively. Note that
  all the  galaxy luminosity functions  listed in this  table are in  units of
  $10^{-2} h^3{\rm  Mpc}^{-3} {\rm d}  \log L$, where  $\log$ is the  10 based
  logarithm. }\label{tab:gax_LF}
\end{deluxetable*}
\end{turnpage}

\begin{turnpage}
\begin{deluxetable*}{cccccccccccc} 
  \tabletypesize{\scriptsize} 
  \tablecaption{The galaxy stellar mass functions $\Phi(M_{\ast})$ }
  \tablewidth{0pt} 
  \tablehead{  &\multicolumn{3}{c}{ALL} && \multicolumn{3}{c}{CENTRAL} && \multicolumn{3}{c}{SATELLITE}\\ 
  \cline{2-4} \cline{6-8} \cline{10-12}\\ (1) & (2) & (3) & (4) & & (5) & (6) &
  (7) && (8) & (9) & (10) \\
   $\log M_{\ast}$ & all & red & blue && all & red & blue && all & red & blue  }

\startdata
  8.2  &  6.7578 $\pm$ 3.5635  &  5.0225 $\pm$ 3.4088  &  1.7354 $\pm$ 1.3697  &&  5.4097 $\pm$ 3.6339  &  3.6743 $\pm$ 3.1606  &  1.7354 $\pm$ 1.3697  &&  1.3481 $\pm$ 1.4459  &  1.3481 $\pm$ 1.4459  &  0.0000 $\pm$ 0.0000 \\
  8.3  &  5.0065 $\pm$ 1.1628  &  2.6774 $\pm$ 0.9420  &  2.3291 $\pm$ 0.8917  &&  2.4939 $\pm$ 0.7368  &  1.3292 $\pm$ 0.6210  &  1.1647 $\pm$ 0.6085  &&  2.5126 $\pm$ 0.7970  &  1.3483 $\pm$ 0.5956  &  1.1644 $\pm$ 0.5844 \\
  8.4  &  5.2026 $\pm$ 0.8331  &  1.8689 $\pm$ 0.5549  &  3.3337 $\pm$ 0.5824  &&  3.7970 $\pm$ 0.8574  &  0.9167 $\pm$ 0.3865  &  2.8803 $\pm$ 0.6688  &&  1.4056 $\pm$ 0.7744  &  0.9522 $\pm$ 0.5715  &  0.4534 $\pm$ 0.3364 \\
  8.5  &  3.6943 $\pm$ 0.6310  &  1.3602 $\pm$ 0.3937  &  2.3340 $\pm$ 0.4160  &&  2.8124 $\pm$ 0.5254  &  0.8612 $\pm$ 0.2399  &  1.9512 $\pm$ 0.4411  &&  0.8819 $\pm$ 0.5271  &  0.4990 $\pm$ 0.3365  &  0.3829 $\pm$ 0.2879 \\
  8.6  &  3.2254 $\pm$ 0.4123  &  0.9867 $\pm$ 0.2701  &  2.2387 $\pm$ 0.3139  &&  2.4063 $\pm$ 0.4036  &  0.5364 $\pm$ 0.1443  &  1.8699 $\pm$ 0.3612  &&  0.8191 $\pm$ 0.3452  &  0.4503 $\pm$ 0.2265  &  0.3687 $\pm$ 0.1858 \\
  8.7  &  3.1947 $\pm$ 0.3411  &  0.8821 $\pm$ 0.2339  &  2.3126 $\pm$ 0.2237  &&  2.0724 $\pm$ 0.3095  &  0.4822 $\pm$ 0.1251  &  1.5902 $\pm$ 0.2715  &&  1.1223 $\pm$ 0.3426  &  0.3999 $\pm$ 0.1989  &  0.7224 $\pm$ 0.2100 \\
  8.8  &  2.7979 $\pm$ 0.2362  &  0.8262 $\pm$ 0.1388  &  1.9717 $\pm$ 0.2255  &&  2.0071 $\pm$ 0.2667  &  0.3576 $\pm$ 0.0824  &  1.6495 $\pm$ 0.2453  &&  0.7908 $\pm$ 0.1654  &  0.4687 $\pm$ 0.1130  &  0.3222 $\pm$ 0.0997 \\
  8.9  &  2.7209 $\pm$ 0.1921  &  0.9936 $\pm$ 0.1432  &  1.7273 $\pm$ 0.1742  &&  1.8136 $\pm$ 0.2374  &  0.4987 $\pm$ 0.0766  &  1.3149 $\pm$ 0.2136  &&  0.9073 $\pm$ 0.1880  &  0.4949 $\pm$ 0.1142  &  0.4124 $\pm$ 0.1215 \\
  9.0  &  2.5366 $\pm$ 0.1544  &  0.7511 $\pm$ 0.1110  &  1.7855 $\pm$ 0.1573  &&  1.7138 $\pm$ 0.1638  &  0.3434 $\pm$ 0.0637  &  1.3704 $\pm$ 0.1687  &&  0.8228 $\pm$ 0.1219  &  0.4076 $\pm$ 0.0788  &  0.4151 $\pm$ 0.0770 \\
  9.1  &  2.8001 $\pm$ 0.1436  &  0.9423 $\pm$ 0.1323  &  1.8578 $\pm$ 0.1762  &&  1.8134 $\pm$ 0.2080  &  0.3792 $\pm$ 0.0744  &  1.4342 $\pm$ 0.1698  &&  0.9867 $\pm$ 0.1670  &  0.5630 $\pm$ 0.1391  &  0.4236 $\pm$ 0.0625 \\
  9.2  &  2.5299 $\pm$ 0.1329  &  0.8579 $\pm$ 0.1015  &  1.6720 $\pm$ 0.1546  &&  1.6551 $\pm$ 0.1690  &  0.3732 $\pm$ 0.0545  &  1.2819 $\pm$ 0.1469  &&  0.8748 $\pm$ 0.1109  &  0.4847 $\pm$ 0.0974  &  0.3901 $\pm$ 0.0431 \\
  9.3  &  2.6046 $\pm$ 0.1105  &  0.8839 $\pm$ 0.0824  &  1.7207 $\pm$ 0.1232  &&  1.7112 $\pm$ 0.1675  &  0.4203 $\pm$ 0.0540  &  1.2908 $\pm$ 0.1388  &&  0.8935 $\pm$ 0.1283  &  0.4636 $\pm$ 0.0881  &  0.4299 $\pm$ 0.0612 \\
  9.4  &  2.6018 $\pm$ 0.1118  &  1.0060 $\pm$ 0.0732  &  1.5958 $\pm$ 0.1207  &&  1.6445 $\pm$ 0.1602  &  0.4455 $\pm$ 0.0551  &  1.1990 $\pm$ 0.1283  &&  0.9573 $\pm$ 0.1105  &  0.5605 $\pm$ 0.0818  &  0.3968 $\pm$ 0.0480 \\
  9.5  &  2.2135 $\pm$ 0.1075  &  0.8682 $\pm$ 0.0492  &  1.3453 $\pm$ 0.1048  &&  1.4016 $\pm$ 0.1165  &  0.4033 $\pm$ 0.0414  &  0.9983 $\pm$ 0.0926  &&  0.8118 $\pm$ 0.0559  &  0.4649 $\pm$ 0.0533  &  0.3470 $\pm$ 0.0317 \\
  9.6  &  2.1244 $\pm$ 0.0969  &  0.9046 $\pm$ 0.0347  &  1.2197 $\pm$ 0.0840  &&  1.4653 $\pm$ 0.1207  &  0.4732 $\pm$ 0.0443  &  0.9921 $\pm$ 0.0951  &&  0.6590 $\pm$ 0.0572  &  0.4314 $\pm$ 0.0407  &  0.2276 $\pm$ 0.0288 \\
  9.7  &  1.9695 $\pm$ 0.0934  &  0.9288 $\pm$ 0.0355  &  1.0407 $\pm$ 0.0762  &&  1.3411 $\pm$ 0.1129  &  0.5037 $\pm$ 0.0502  &  0.8374 $\pm$ 0.0763  &&  0.6284 $\pm$ 0.0531  &  0.4251 $\pm$ 0.0434  &  0.2033 $\pm$ 0.0194 \\
  9.8  &  1.7968 $\pm$ 0.0687  &  0.8633 $\pm$ 0.0252  &  0.9335 $\pm$ 0.0614  &&  1.2160 $\pm$ 0.0882  &  0.4676 $\pm$ 0.0395  &  0.7483 $\pm$ 0.0616  &&  0.5808 $\pm$ 0.0451  &  0.3956 $\pm$ 0.0395  &  0.1852 $\pm$ 0.0144 \\
  9.9  &  1.7543 $\pm$ 0.0808  &  0.9275 $\pm$ 0.0410  &  0.8268 $\pm$ 0.0538  &&  1.1954 $\pm$ 0.0921  &  0.5415 $\pm$ 0.0491  &  0.6539 $\pm$ 0.0545  &&  0.5589 $\pm$ 0.0340  &  0.3859 $\pm$ 0.0271  &  0.1729 $\pm$ 0.0134 \\
 10.0  &  1.6529 $\pm$ 0.0658  &  0.9104 $\pm$ 0.0335  &  0.7426 $\pm$ 0.0427  &&  1.1378 $\pm$ 0.0845  &  0.5414 $\pm$ 0.0472  &  0.5963 $\pm$ 0.0451  &&  0.5152 $\pm$ 0.0338  &  0.3689 $\pm$ 0.0283  &  0.1462 $\pm$ 0.0115 \\
 10.1  &  1.6162 $\pm$ 0.0660  &  0.9446 $\pm$ 0.0356  &  0.6716 $\pm$ 0.0398  &&  1.1427 $\pm$ 0.0837  &  0.5939 $\pm$ 0.0483  &  0.5488 $\pm$ 0.0431  &&  0.4735 $\pm$ 0.0316  &  0.3507 $\pm$ 0.0264  &  0.1228 $\pm$ 0.0100 \\
 10.2  &  1.5700 $\pm$ 0.0631  &  0.9791 $\pm$ 0.0388  &  0.5908 $\pm$ 0.0340  &&  1.1120 $\pm$ 0.0814  &  0.6402 $\pm$ 0.0531  &  0.4718 $\pm$ 0.0351  &&  0.4580 $\pm$ 0.0310  &  0.3389 $\pm$ 0.0276  &  0.1190 $\pm$ 0.0076 \\
 10.3  &  1.5025 $\pm$ 0.0564  &  0.9766 $\pm$ 0.0325  &  0.5259 $\pm$ 0.0298  &&  1.0918 $\pm$ 0.0766  &  0.6635 $\pm$ 0.0503  &  0.4284 $\pm$ 0.0320  &&  0.4107 $\pm$ 0.0308  &  0.3132 $\pm$ 0.0271  &  0.0975 $\pm$ 0.0066 \\
 10.4  &  1.3108 $\pm$ 0.0480  &  0.8909 $\pm$ 0.0302  &  0.4199 $\pm$ 0.0230  &&  0.9663 $\pm$ 0.0682  &  0.6177 $\pm$ 0.0479  &  0.3486 $\pm$ 0.0252  &&  0.3445 $\pm$ 0.0285  &  0.2732 $\pm$ 0.0247  &  0.0713 $\pm$ 0.0057 \\
 10.5  &  1.0826 $\pm$ 0.0356  &  0.7608 $\pm$ 0.0245  &  0.3218 $\pm$ 0.0155  &&  0.8213 $\pm$ 0.0529  &  0.5522 $\pm$ 0.0384  &  0.2691 $\pm$ 0.0186  &&  0.2613 $\pm$ 0.0237  &  0.2087 $\pm$ 0.0197  &  0.0527 $\pm$ 0.0057 \\
 10.6  &  0.8499 $\pm$ 0.0267  &  0.6080 $\pm$ 0.0188  &  0.2419 $\pm$ 0.0111  &&  0.6604 $\pm$ 0.0418  &  0.4549 $\pm$ 0.0309  &  0.2055 $\pm$ 0.0138  &&  0.1896 $\pm$ 0.0193  &  0.1532 $\pm$ 0.0159  &  0.0364 $\pm$ 0.0046 \\
 10.7  &  0.6440 $\pm$ 0.0193  &  0.4772 $\pm$ 0.0145  &  0.1667 $\pm$ 0.0074  &&  0.5182 $\pm$ 0.0311  &  0.3753 $\pm$ 0.0242  &  0.1429 $\pm$ 0.0092  &&  0.1257 $\pm$ 0.0151  &  0.1019 $\pm$ 0.0129  &  0.0238 $\pm$ 0.0032 \\
 10.8  &  0.4582 $\pm$ 0.0130  &  0.3556 $\pm$ 0.0107  &  0.1026 $\pm$ 0.0039  &&  0.3811 $\pm$ 0.0211  &  0.2910 $\pm$ 0.0174  &  0.0901 $\pm$ 0.0051  &&  0.0771 $\pm$ 0.0101  &  0.0646 $\pm$ 0.0087  &  0.0125 $\pm$ 0.0020 \\
 10.9  &  0.3009 $\pm$ 0.0071  &  0.2436 $\pm$ 0.0069  &  0.0573 $\pm$ 0.0014  &&  0.2572 $\pm$ 0.0123  &  0.2064 $\pm$ 0.0114  &  0.0508 $\pm$ 0.0019  &&  0.0437 $\pm$ 0.0067  &  0.0373 $\pm$ 0.0059  &  0.0064 $\pm$ 0.0012 \\
 11.0  &  0.1846 $\pm$ 0.0047  &  0.1573 $\pm$ 0.0048  &  0.0273 $\pm$ 0.0011  &&  0.1627 $\pm$ 0.0073  &  0.1379 $\pm$ 0.0071  &  0.0248 $\pm$ 0.0008  &&  0.0219 $\pm$ 0.0035  &  0.0194 $\pm$ 0.0031  &  0.0025 $\pm$ 0.0006 \\
 11.1  &  0.1066 $\pm$ 0.0022  &  0.0931 $\pm$ 0.0027  &  0.0135 $\pm$ 0.0012  &&  0.0964 $\pm$ 0.0037  &  0.0839 $\pm$ 0.0039  &  0.0125 $\pm$ 0.0008  &&  0.0102 $\pm$ 0.0020  &  0.0092 $\pm$ 0.0016  &  0.0010 $\pm$ 0.0005 \\
 11.2  &  0.0575 $\pm$ 0.0011  &  0.0520 $\pm$ 0.0017  &  0.0055 $\pm$ 0.0009  &&  0.0531 $\pm$ 0.0018  &  0.0480 $\pm$ 0.0023  &  0.0052 $\pm$ 0.0007  &&  0.0044 $\pm$ 0.0010  &  0.0040 $\pm$ 0.0008  &  0.0003 $\pm$ 0.0002 \\
 11.3  &  0.0294 $\pm$ 0.0006  &  0.0271 $\pm$ 0.0009  &  0.0023 $\pm$ 0.0004  &&  0.0281 $\pm$ 0.0009  &  0.0258 $\pm$ 0.0011  &  0.0022 $\pm$ 0.0004  &&  0.0014 $\pm$ 0.0004  &  0.0013 $\pm$ 0.0003  &  0.0001 $\pm$ 0.0001 \\
 11.4  &  0.0131 $\pm$ 0.0004  &  0.0122 $\pm$ 0.0005  &  0.0010 $\pm$ 0.0002  &&  0.0127 $\pm$ 0.0005  &  0.0117 $\pm$ 0.0005  &  0.0009 $\pm$ 0.0002  &&  0.0004 $\pm$ 0.0001  &  0.0004 $\pm$ 0.0001  &  0.0000 $\pm$ 0.0000 \\
 11.5  &  0.0047 $\pm$ 0.0003  &  0.0044 $\pm$ 0.0003  &  0.0003 $\pm$ 0.0001  &&  0.0047 $\pm$ 0.0003  &  0.0044 $\pm$ 0.0003  &  0.0003 $\pm$ 0.0001  &&  0.0000 $\pm$ 0.0000  &  0.0000 $\pm$ 0.0000  &  0.0000 $\pm$ 0.0000 \\
 11.6  &  0.0012 $\pm$ 0.0001  &  0.0011 $\pm$ 0.0001  &  0.0001 $\pm$ 0.0000  &&  0.0012 $\pm$ 0.0001  &  0.0011 $\pm$ 0.0001  &  0.0001 $\pm$ 0.0000  &&  0.0000 $\pm$ 0.0000  &  0.0000 $\pm$ 0.0000  &  0.0000 $\pm$ 0.0000 
\enddata
 
\tablecomments{ Column (1): the median  of the logarithm of the galaxy stellar
  mass with  bin width  $\Delta \log M_{\ast}  = 0.05$.   Column (2 -  4): the
  stellar mass functions of all, red  and blue for 'ALL' group members. Column
  (5 - 7): the stellar mass functions of all, red and blue for 'CENTRAL' group
  members.  Column (8  - 10): the stellar mass functions of  all, red and blue
  for  'SATELLITE' group  members.   Note  that all  the  galaxy stellar  mass
  functions listed  in this table are  in units of  $10^{-2} h^3{\rm Mpc}^{-3}
  {\rm    d}   \log    M_{\ast}$,    where   $\log$    is    the   10    based
  logarithm. }\label{tab:gax_MF}
\end{deluxetable*}
\end{turnpage}

\begin{figure*} \plotone{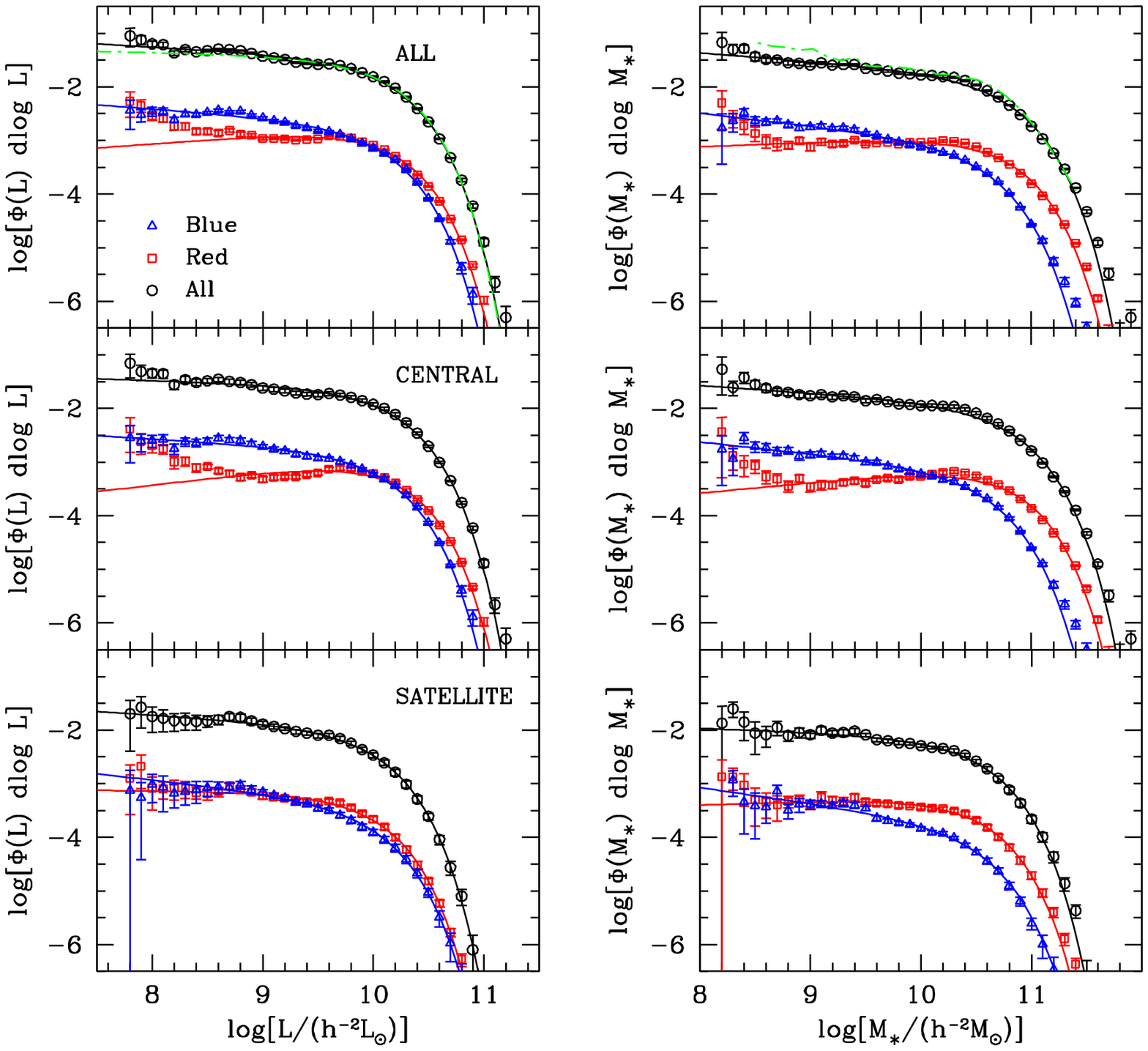}
  \caption{The  galaxy luminosity  functions  (left panels)  and stellar  mass
    functions (right panels).  The upper, middle and lower panels show results
    obtained from  all, central and satellite group  members, respectively. In
    each panel,  the open circles,  squares and triangles with  error-bars are
    the luminosity  or stellar mass functions  of all, red  and blue galaxies,
    respectively,  where  the  errors  are  obtained from  the  200  bootstrap
    samples. Note that  the results for red and blue  galaxies are scaled down
    by a  factor of 10, for  clarity.  The solid  lines in each panel  are the
    best fitting Schechter functions.   For comparison, the dot-dashed line in
    the  upper-left panel  is the  best  fit luminosity  function obtained  by
    Blanton et al.   (2003b), while that in the  upper-right panel corresponds
    to the stellar mass function obtained by Bell et al.  (2003) from the SDSS
    Early Data Release.}
\label{fig:gax_LF_MF}
\end{figure*}

\begin{deluxetable}{llccc}
  \tabletypesize{\scriptsize} \tablecaption{The best fit parameters for the
    galaxy luminosity functions and stellar mass functions} \tablewidth{0pt}
  \tablehead{Member type & color & $\phi^{\star}$ & $\alpha$ & $\log L^{\star}$ \\
    \cline{1-5}\\ (1) & (2) & (3) & (4) & (5) }

\startdata
ALL       & all       & 0.03167 & -1.117 & 10.095 \\
--        & red       & 0.01810 & -0.846 & 10.080 \\
--        & blue      & 0.01890 & -1.154 & 10.026 \\
CENTRAL   & all       & 0.02370 & -1.069 & 10.114 \\
--        & red       & 0.01223 & -0.755 & 10.106 \\
--        & blue      & 0.01523 & -1.123 & 10.033 \\
SATELLITE & all       & 0.01051 & -1.134 &  9.947 \\
--        & red       & 0.00682 & -1.018 &  9.923 \\
--        & blue      & 0.00410 & -1.235 &  9.951 \\
\cline{1-5} \\
Member type & color & $\phi^{\star}$ & $\alpha$ & $\log M^{\star}$ \\
\cline{1-5}\\
ALL       & all       & 0.01546 & -1.164 & 10.717 \\
--        & red       & 0.01285 & -0.916 & 10.691 \\
--        & blue      & 0.00836 & -1.233 & 10.508 \\
CENTRAL   & all       & 0.01084 & -1.143 & 10.758 \\
--        & red       & 0.00903 & -0.803 & 10.717 \\
--        & blue      & 0.00668 & -1.218 & 10.522 \\
SATELLITE & all       & 0.00692 & -1.078 & 10.483 \\
--        & red       & 0.00589 & -0.932 & 10.457 \\
--        & blue      & 0.00165 & -1.290 & 10.450 
\enddata
 
\tablecomments{  Column  (1):  the  member  type. Column  (2):  the  color  of
  galaxies. Column (3-5): the best fit parameters for the luminosity functions
  (upper  part)   and  stellar  mass   functions  (lower  part).    Note  that
  $\phi^{\star}$  listed  in column  3  are  presented  in terms  of  $h^3{\rm
    Mpc}^{-3} {\rm d} \log L$  (or $h^3{\rm Mpc}^{-3} {\rm d} \log M_{\ast}$),
  where $\log$ is the 10 based logarithm.  }\label{tab:fit}
\end{deluxetable}

In  this section we  estimate the  luminosity and  stellar mass  functions for
different  populations  of galaxies.   Both  functions  have been  extensively
investigated  in   the  literature  for  galaxies  of   different  colors  and
morphological types (e.g.,  Lin et al. 1996; Norberg et  al. 2002; Madgwick et
al. 2002; Blanton et al. 2003b; Bell  et al. 2003; Fontana et al. 2006).  Note
that in a very recent paper, based on the group catalogue constructed from the
2 degree Field  Galaxy Redshift Survey (2dFGRS; Colless et  al., 2001) by Tago
et al. (2006), Tempel et  al. (2008) measured luminosity functions for various
contents of  galaxies (e.g., central,  second ranked, satellite,  isolated) in
groups, as well as for groups. Here we focus on the difference between central
and satellite  galaxies.  We  use direct counting  to estimate  the luminosity
function.   For each  galaxy at  a given  redshift we  calculate  its absolute
magnitude  limit   according  to  Eq.\,(\ref{eq:maglim}).    If  the  absolute
magnitude of  the galaxy is  fainter than this  limit, it is removed  from the
counting list.  If  the galaxy is not removed, we  first calculate the maximum
redshift at  which the galaxy (with  its absolute magnitude)  can be observed.
We then  calculate the  comoving volume, $V_{\rm  com}$, between  this maximum
redshift  and a minimum  redshift $z=0.01$.   In the  counting, the  galaxy is
assigned a weight,
\begin{equation}\label{eq:weight} w_i=\frac 1 {V_{\rm com}\calC}\,
\end{equation}
where  $\calC$  is the  redshift  completeness  factor  in the  NYU-VAGC.   We
calculate  the   luminosity  functions  for  all,  red,   and  blue  galaxies,
respectively.  The corresponding results are  shown in the upper-left panel of
Fig. \ref{fig:gax_LF_MF},  where open circles,  squares and triangles  are the
results for all, red and blue galaxies, respectively.  For clarity the results
for  red and  blue  galaxies are  shifted downwards  by  a factor  of 10.   By
comparing the  LFs for  red and blue  galaxies, one  sees that there  are more
(fewer) red  galaxies than  blue ones at  luminosities $\log L\ga  9.8$ ($\log
L\la 9.8$).  For very faint galaxies  with $\log L\la 9.0$, the red population
increase dramatically, exceeding that of blue galaxies at $\log L\sim 8.0$ .

We can further separate the galaxies into centrals and satellites according to
their memberships  in the groups.  The corresponding  luminosity functions for
all, red and blue galaxies are  shown in the middle-left and lower-left panels
of  Fig.   \ref{fig:gax_LF_MF}, respectively.   The  color  dependence of  the
luminosity function for the centrals resembles that of the overall population.
However,  for  the satellite  population,  the  color  dependence is  somewhat
different, especially  around $L  = 10^9 h^{-2}  \Lsun$, where  the luminosity
function is not suppressed relative to that of blue satellites, as is the case
for the centrals.   One interesting feature in the  luminosity function of red
central  galaxies is  that  there are  many  very faint  red central  galaxies
(slightly more than  the blue ones) with luminosity $\log  L\sim 8.0$.  Such a
population is not expected in the standard galaxy formation models, where very
small halos  are expected to  host only blue  centrals.  However, Wang,  Mo \&
Jing (2007)  found that the  large-scale tidal field may  effectively truncate
the mass  accretion into small halos.   If gas accretion is  also truncated in
this process,  the central galaxies  in these small  halos are expected  to be
red.  More recently, Ludlow et al.  (2008) found that some sub-halos that have
at some  time been within the virial  radius of their main  progenitors can be
ejected,  so  that  some  low-mass  halos at  $z=0$  outside/near  the  larger
virialized  halos may have  experienced tidal  and ram-pressure  stripping, so
that they have stopped forming stars. In both scenarios, a population of faint
red galaxies is expected to be present in the vicinity of high density regions
but outside large,  virialized halos.  This population may  be responsible for
the upturn  of the luminosity  function of red  central galaxies at  the faint
end. In a separate paper (Wang  et al.  2008b), we will discuss the properties
and spatial clustering  of the faint red population in  more detail, and check
the contamination due to false groups  near massive ones using mock galaxy and
group catalogs.

\begin{figure*} \plotone{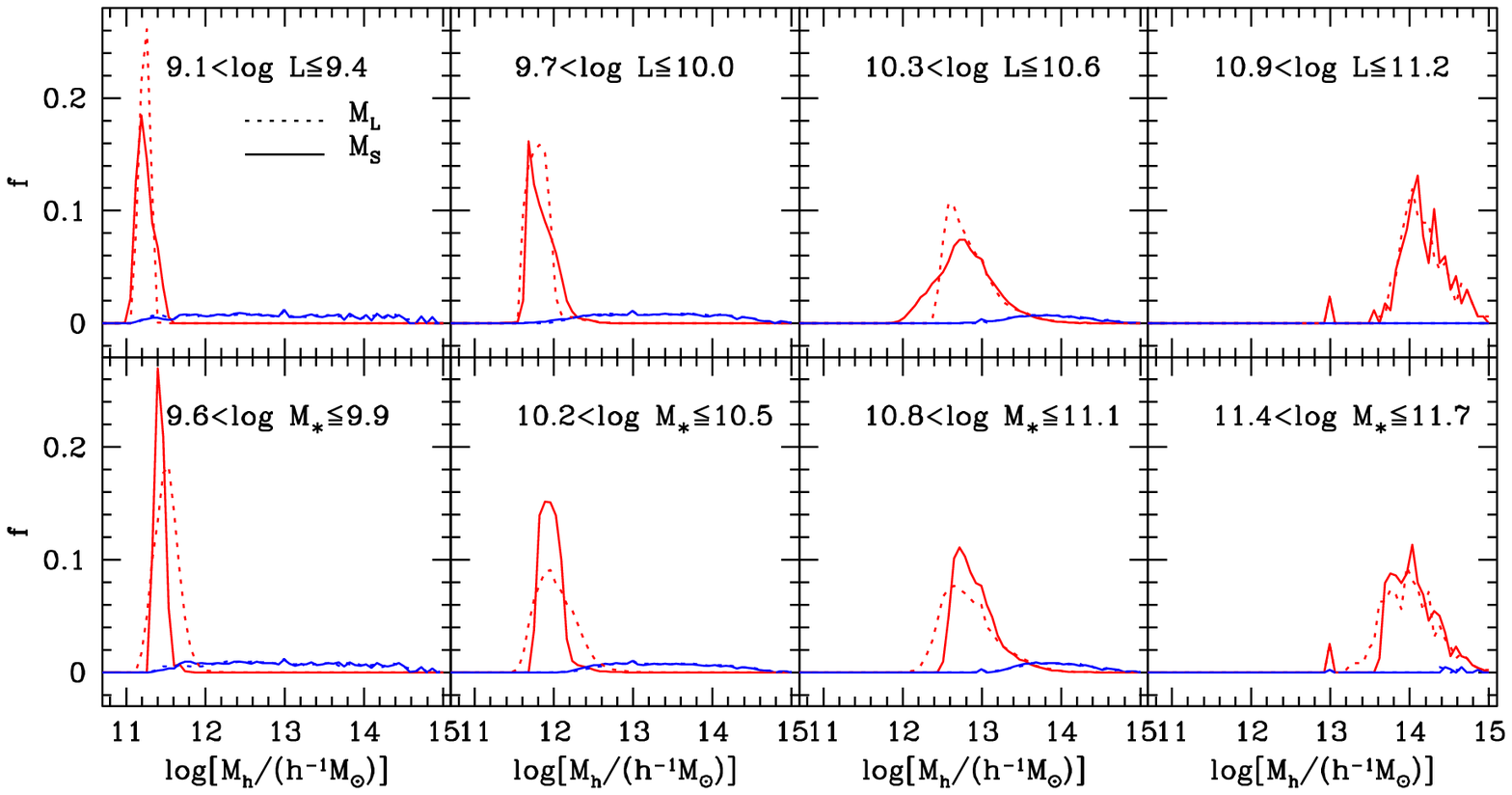}
  \caption{The  host  halo mass  distributions  for  central  (red lines)  and
    satellite (blue  lines) galaxies.  Shown  in the upper panels  are results
    for galaxies in different luminosity bins as indicated. Shown in the lower
    panels  are  results  for  galaxies  in different  stellar  mass  bins  as
    indicated.  The dotted  and solid  lines  represent the  results for  halo
    masses, $M_L$ and $M_S$, respectively.  }
\label{fig:f_L_M}
\end{figure*}

We  also measure the  stellar mass  functions separately  for all,  red, blue,
central  and  satellite  galaxies,   taking  into  account  the  stellar  mass
completeness  limit described  by Eq.(\ref{eq:mstarlim}).  Only  galaxies with
stellar masses above the completeness limit  are used in the estimate.  In the
right panels of  Fig. \ref{fig:gax_LF_MF}, we show the  stellar mass functions
using the  same symbols as  in the left  panels. The general behaviors  of the
stellar  mass  functions are  very  similar  to  the corresponding  luminosity
functions.   For  reference,  we   list  the  luminosity  functions  in  Table
\ref{tab:gax_LF}, and the stellar mass functions in Table \ref{tab:gax_MF}.

We use the following Schechter function to fit the luminosity function:
\begin{equation}\label{eq:phi_L} 
 \Phi(L)   =   \phi^{\star}\left  (   {L\over
L^{\star}}\right )^{(\alpha+1)} {\rm exp} \left[- {L\over L^{\star}}\right ]\,.
\end{equation}
For the stellar mass function luminosity functions, we use a similar 
model:
\begin{equation}\label{eq:phi_M}
\Phi(M_{\ast}) = \phi^{\star}\left ( {M_{\ast}\over M^{\star}}\right )^{(\alpha+1)}
  {\rm exp} \left[-  {M_{\ast}\over M^{\star}}\right ]\,.
\end{equation}
For each model, there are three free parameters, the amplitude $\phi^{\star}$,
the faint end slope $\alpha$ and the characteristic luminosity $L^{\star}$ (or
stellar mass  $M^{\star}$). Using the  least $\chi^2$ fitting to  the measured
luminosity and  stellar mass functions  shown in Fig.  \ref{fig:gax_LF_MF}, we
find  the  best  fit  values  of  $\phi^{\star}$,  $\alpha$,  $L^{\star}$  (or
$M^{\star}$), which are  listed in Table \ref{tab:fit}.  The  best fit results
are shown as the solid lines in Fig.\,\ref{fig:gax_LF_MF}.

Comparing  the data  with  the best  fit,  one sees  that  the Schechter  form
describes the data remarkably well over the luminosity range $\log L \ga 9.0$.
At  the fainter  end, however,  the data  reveals an  upturn, which  is almost
entirely due  to the  red centrals.   The stellar mass  functions also  show a
steepening at the  low mass end, which again owes mainly  to the population of
red centrals.

The galaxy luminosity and stellar mass functions have been estimated before by
various authors.  As  an illustration, we show the best  fit of the luminosity
function  obtained  by Blanton  et  al.   (2003b) from  the  SDSS  DR2 as  the
dot-dashed  line in  the upper-left  panel of  Fig.   \ref{fig:gax_LF_MF}.  As
expected, our  measurements are in  excellent agreement with theirs.   For the
stellar mass  function, we show as the  dot-dashed line the result  of Bell et
al. (2003)  obtained from the SDSS  Early Data Release (EDR;  Stoughton et al.
2002).   The agreement  with  our  measurement is  remarkably  good for  $\log
M_{\ast} \ga 9.5$.   For lower masses, however, the  mass function obtained by
Bell  et al.  (2003) is  significantly  higher.  The  discrepancy most  likely
results  from the  different  data samples  (DR4  vs.  EDR)  used  in the  two
analyses.  For   low-mass  galaxies,  the  cosmic   variance  is  significant,
especially in EDR  because only a few hundred galaxies in  a small volume were
used to  measure the stellar  mass function. In  addition, we have  taken into
account  the redshift completeness  of galaxies  using the  completeness masks
provided by the NYU team, and the stellar mass limit is treated more carefully
in our analysis using Eq. \ref{eq:mstarlim}.  The behavior of the stellar mass
function  in  the  low-mass  end  has  also been  investigated  by  Baldry  et
al. (2004,  2008) and Panter et  al. (2007).  Unfortunately,  the situation is
still unclear, partly because of the limited sample volume, and partly because
of the uncertainties in the luminosity-mass conversion.

To see where galaxies of different luminosities and stellar masses are hosted,
we plot  in Fig. \ref{fig:f_L_M} the  host halo mass  distribution for central
(red  lines)  and satellite  (blue  lines)  galaxies.  Results are  shown  for
galaxies  in  different luminosity  (upper  panels)  and  stellar mass  (lower
panels)  bins, as  indicated \footnote{Note  that in  the group  catalog, halo
masses are provided  only for groups whose central  galaxies are brighter than
$\rmag=-19.5$.   For groups  with a  fainter central  galaxy we  use  the mean
mass-to-light      (Eq.\ref{eq:Lc_fit})      and     halo-to-stellar      mass
(Eq.\ref{eq:Mc_fit})  ratios obtained in  Section \ref{sec_small}  to estimate
their halo masses.}.  Almost all  bright (or massive) galaxies are centrals in
massive halos.  About  3/4 of faint (low-mass) galaxies  are centrals in small
halos with a very narrow mass distribution, and the rest 1/4 are satellites in
halos that  cover a wide range in  mass. The results are  shown separately for
halo masses, $M_L$ and $M_S$.  Although the results for satellite galaxies and
the mean  halo mass  for central galaxies  are similar  for the two  halo mass
estimates, the widths  of the halo mass distribution  for central galaxies are
quite  different.  This  is caused  by the  fact that  these is  some spurious
correlation between $L_{19.5}$ ($M_{\rm stellar}$) and the luminosity (stellar
mass)  of  the  central  galaxy,  especially  for  low-mass  halos  where  the
luminosity  (stellar) content  is dominated  by the  central galaxy.   We have
measured the  halo mass distribution from both  samples II and III,  and we do
not find  any significant difference  between the results. Therefore  only the
results for sample II are plotted in Fig.\ref{fig:f_L_M}. The general behavior
of  the  halo mass  distribution  of the  central  and  satellite galaxies  is
consistent  with that  predicted  by the  CLF  and HOD  models  (e.g. Yang  et
al. 2003; Zheng et al. 2005).
 
\section{The conditional stellar mass function} 
\label{sec_CSMFs}

\begin{figure*} \plotone{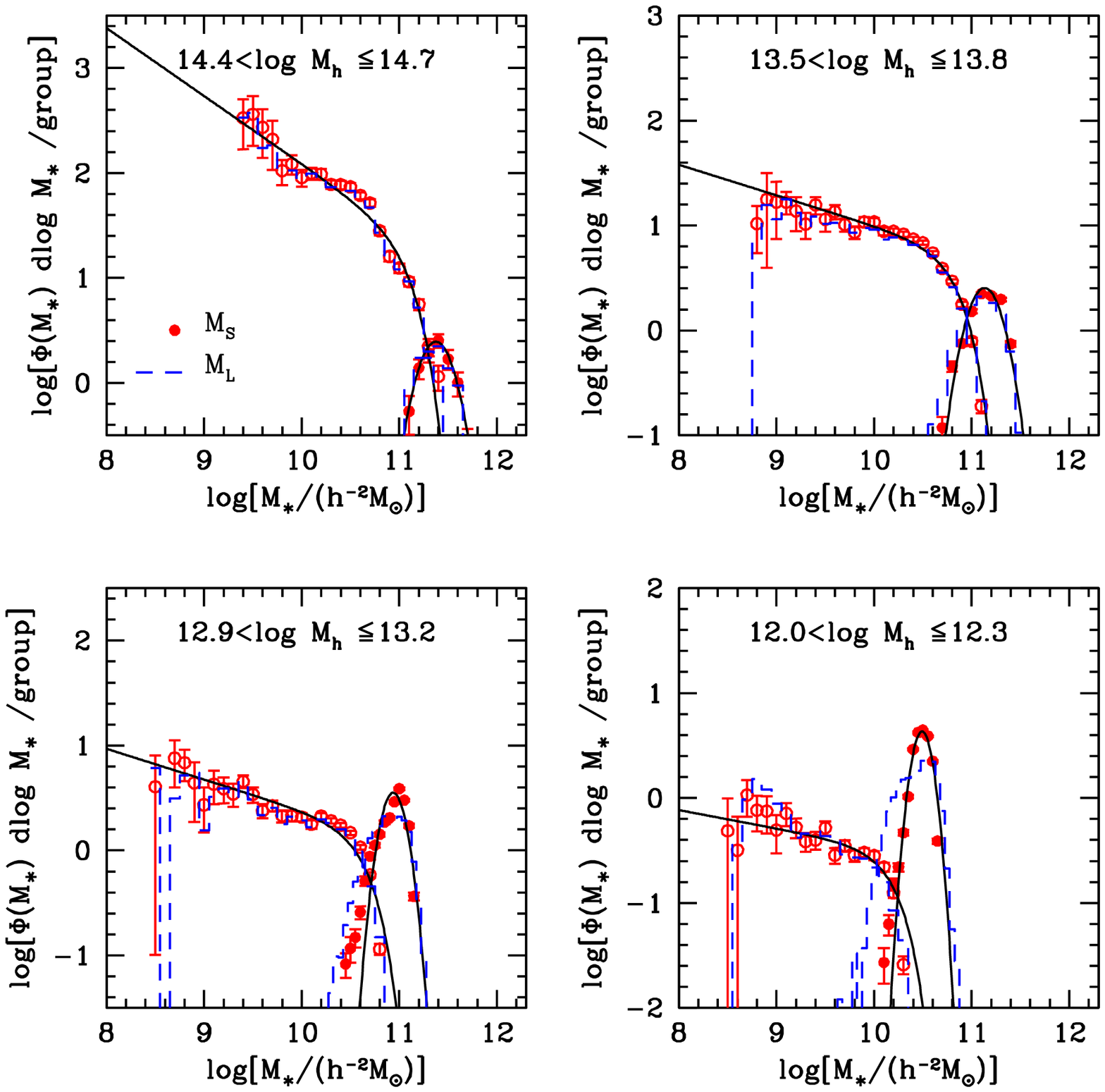}
  \caption{The  conditional  stellar mass  functions  (CSMFs)  of galaxies  in
    groups of different  mass bins.  Symbols correspond to  the CSMFs obtained
    using  $M_S$ as  halo  mass (estimated  according  to the  ranking of  the
    characteristic  group  stellar  masses),   with  solid  and  open  circles
    indicating  the   contributions  from  central   and  satellite  galaxies,
    respectively.  The error-bars reflect the 1-$\sigma$ scatter obtained from
    200 bootstrap samples for $M_L$  and $M_S$, respectively.  The solid lines
    indicate the  best-fit parameterizations (equations~[\ref{eq:CSMF_fit}] to
    [\ref{eq:phi_s}]).  For  comparison, we also show, with  dashed lines, the
    CSMFs  obtained using  $M_L$  as  halo mass  (estimated  according to  the
    ranking of the characteristic group luminosity).}
\label{fig:CSMF}
\end{figure*}
\begin{figure*} \plotone{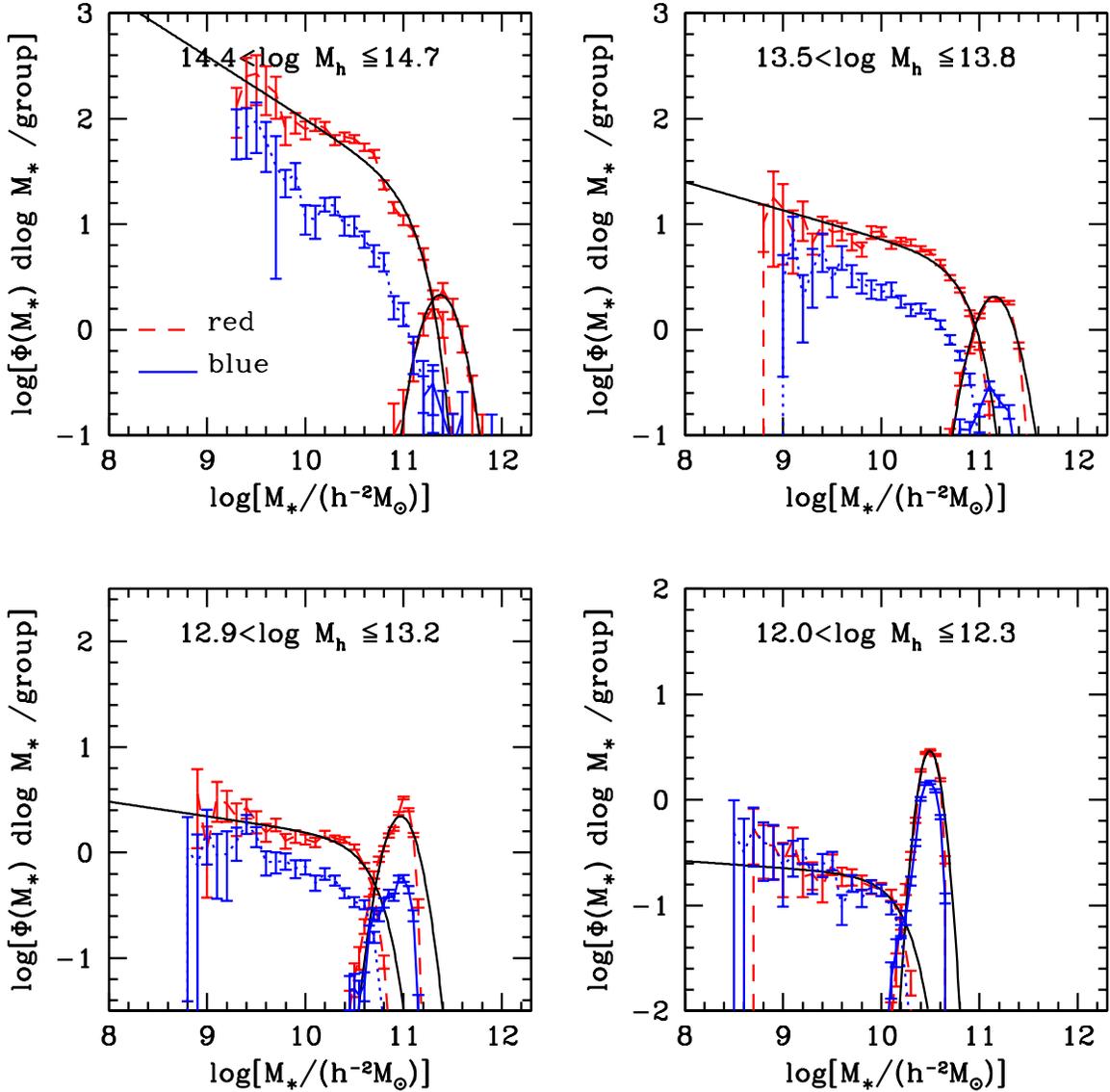}
  \caption{Similar to Fig.~\ref{fig:CSMF}, but here  we show the CSMFs for red
    (dashed lines) and blue (dotted lines) galaxies. In both cases the central
    and  satellite components  of  the CSMFs  are  indicated separately.   The
    error-bars, again obtained using 200  bootstrap samples, are shown for the
    red and blue galaxies, separately.   The solid lines indicate the best-fit
    parameterizations (equations~[\ref{eq:CSMF_fit}]  to [\ref{eq:phi_s}]) for
    red galaxies. Results shown are for halo masses $M_S$ only. }
\label{fig:CSMF_color}
\end{figure*}
\begin{figure*}\plotone{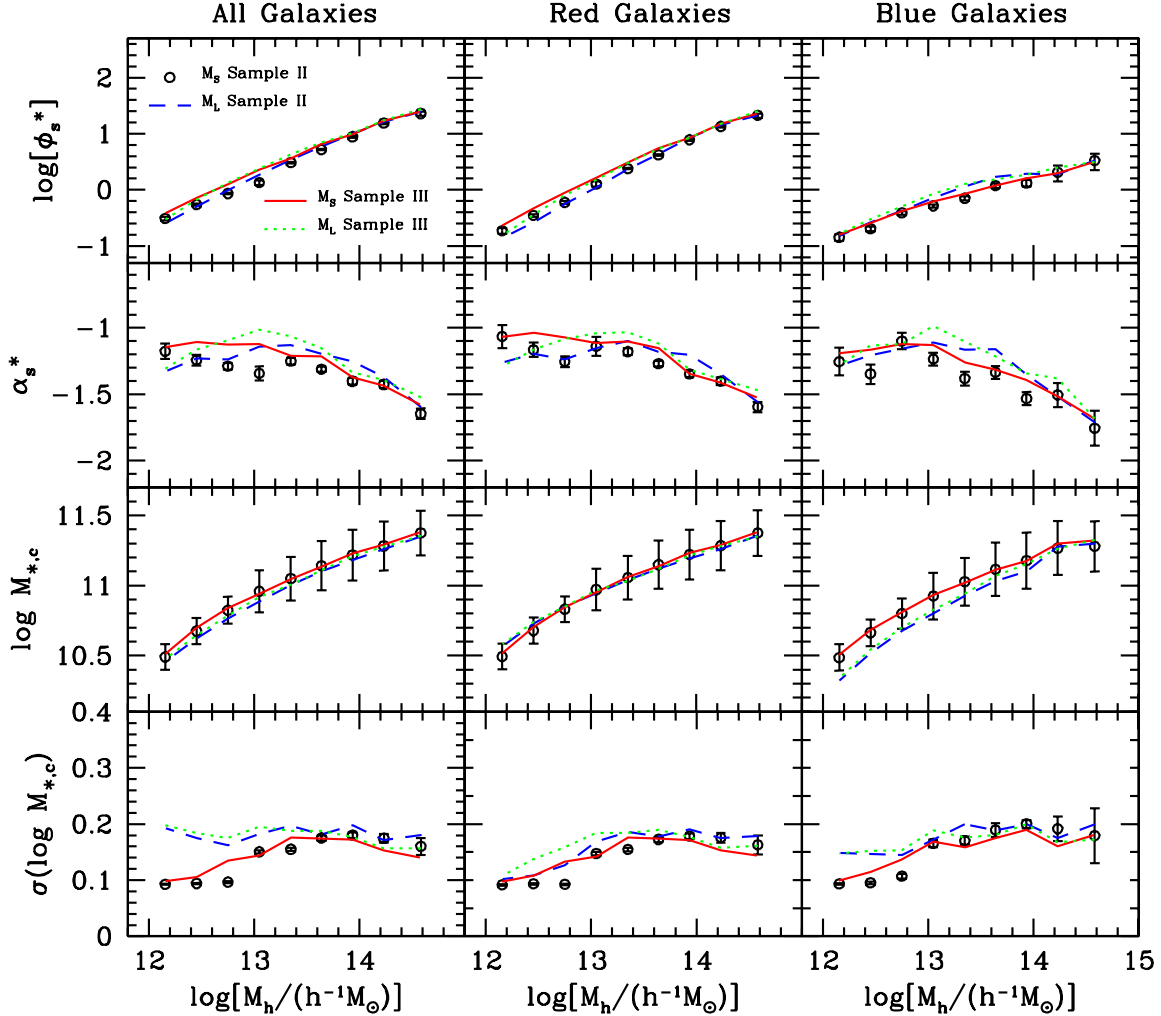}
  \caption{The    best   fit    parameters   ($\phi_s^{\star}$    upper   row,
    $\alpha_s^{\star}$  second  row, $M_{\ast,c}$  third  row, and  $\sigma_c$
    bottom   row)   to   the   CSMFs   shown   in   Figs.~\ref{fig:CSMF}   and
    ~\ref{fig:CSMF_color}, as functions of halo  mass.  Panels on the left, in
    the middle, and on the right show results for all, red, and blue galaxies,
    respectively. Since we have two  different halo mass estimators ($M_S$ and
    $M_L$) and two main group samples  (II and III), we have obtained CLFs for
    four  different combinations  of  sample and  group  mass estimator.   The
    results for  all four combinations  are shown using different  symbols and
    line-styles, as indicated.  The error-bars in the first  two and last rows
    indicate the 1-$\sigma$ variances obtained from our 200 bootstrap samples.
    In  the third row  of panels,  however, the  error-bars correspond  to the
    log-normal scatter,  $\sigma_c$, shown in  the bottom row of  panels.  For
    clarity the error-bars are only  shown for the `$M_S$-sample II' case, but
    they are very similar for the other four cases.  }
\label{fig:fit_CSMF}
\end{figure*}
\begin{deluxetable*}{ccccccc}
  \tabletypesize{\scriptsize} 
  \tablecaption{The best fit parameters of the CSMFs for all, red and blue galaxies} 
  \tablewidth{0pt} 
  \tablehead{ Galaxy type & $\log [M_{h}]$ & $\log \langle [M_{h}]\rangle$ &
  $\phi^*_s$ & $\alpha^*_s$ & $\log M_{\ast,c}$ & $\sigma_c$ \\ 
  \cline{1-7}\\ (1) & (2) & (3) & (4) & (5) & (6) & (7)} 
  
\startdata

& [14.40, 15.00) & $ 14.58$ & $ 24.91\pm  2.51$ & $ -1.59\pm  0.05$ & $11.364\pm 0.018$ & $ 0.159\pm 0.017$\\
& [14.10, 14.40) & $ 14.23$ & $ 16.49\pm  0.79$ & $ -1.40\pm  0.03$ & $11.277\pm 0.013$ & $ 0.164\pm 0.011$\\
& [13.80, 14.10) & $ 13.94$ & $  9.72\pm  0.63$ & $ -1.34\pm  0.07$ & $11.209\pm 0.020$ & $ 0.182\pm 0.011$\\
& [13.50, 13.80) & $ 13.64$ & $  6.15\pm  0.77$ & $ -1.22\pm  0.07$ & $11.122\pm 0.020$ & $ 0.180\pm 0.007$\\
ALL & [13.20, 13.50) & $ 13.35$ & $  3.61\pm  0.52$ & $ -1.16\pm  0.08$ & $11.026\pm 0.024$ & $ 0.179\pm 0.018$\\
& [12.90, 13.20) & $ 13.05$ & $  1.96\pm  0.47$ & $ -1.15\pm  0.14$ & $10.926\pm 0.032$ & $ 0.168\pm 0.025$\\
& [12.60, 12.90) & $ 12.75$ & $  1.10\pm  0.22$ & $ -1.19\pm  0.09$ & $10.803\pm 0.034$ & $ 0.142\pm 0.035$\\
& [12.30, 12.60) & $ 12.45$ & $  0.61\pm  0.10$ & $ -1.19\pm  0.06$ & $10.660\pm 0.033$ & $ 0.140\pm 0.046$\\
& [12.00, 12.30) & $ 12.16$ & $  0.31\pm  0.06$ & $ -1.24\pm  0.09$ & $10.485\pm 0.022$ & $ 0.145\pm 0.058$\\
\cline{1-7}\\
& [14.40, 15.00) & $ 14.58$ & $ 22.64\pm  2.66$ & $ -1.54\pm  0.05$ & $11.365\pm 0.020$ & $ 0.162\pm 0.018$\\
& [14.10, 14.40) & $ 14.23$ & $ 14.40\pm  0.74$ & $ -1.39\pm  0.03$ & $11.278\pm 0.013$ & $ 0.165\pm 0.012$\\
& [13.80, 14.10) & $ 13.94$ & $  8.30\pm  0.33$ & $ -1.30\pm  0.07$ & $11.216\pm 0.017$ & $ 0.179\pm 0.008$\\
& [13.50, 13.80) & $ 13.64$ & $  4.86\pm  0.73$ & $ -1.18\pm  0.06$ & $11.131\pm 0.015$ & $ 0.179\pm 0.008$\\
RED & [13.20, 13.50) & $ 13.35$ & $  2.71\pm  0.39$ & $ -1.10\pm  0.06$ & $11.047\pm 0.014$ & $ 0.175\pm 0.015$\\
& [12.90, 13.20) & $ 13.05$ & $  1.37\pm  0.24$ & $ -1.11\pm  0.07$ & $10.958\pm 0.012$ & $ 0.161\pm 0.020$\\
& [12.60, 12.90) & $ 12.75$ & $  0.71\pm  0.15$ & $ -1.16\pm  0.10$ & $10.844\pm 0.009$ & $ 0.127\pm 0.027$\\
& [12.30, 12.60) & $ 12.45$ & $  0.36\pm  0.09$ & $ -1.14\pm  0.07$ & $10.713\pm 0.027$ & $ 0.112\pm 0.018$\\
& [12.00, 12.30) & $ 12.16$ & $  0.18\pm  0.04$ & $ -1.17\pm  0.12$ & $10.539\pm 0.042$ & $ 0.100\pm 0.007$\\
\cline{1-7}\\
& [14.40, 15.00) & $ 14.58$ & $  3.23\pm  5.24$ & $ -1.71\pm  0.31$ & $11.305\pm 0.134$ & $ 0.183\pm 0.049$\\
& [14.10, 14.40) & $ 14.23$ & $  2.11\pm  1.06$ & $ -1.48\pm  0.18$ & $11.276\pm 0.073$ & $ 0.174\pm 0.044$\\
& [13.80, 14.10) & $ 13.94$ & $  1.67\pm  0.27$ & $ -1.41\pm  0.09$ & $11.152\pm 0.035$ & $ 0.197\pm 0.013$\\
& [13.50, 13.80) & $ 13.64$ & $  1.41\pm  0.26$ & $ -1.26\pm  0.08$ & $11.082\pm 0.040$ & $ 0.183\pm 0.012$\\
BLUE & [13.20, 13.50) & $ 13.35$ & $  0.99\pm  0.26$ & $ -1.23\pm  0.12$ & $10.978\pm 0.053$ & $ 0.176\pm 0.017$\\
& [12.90, 13.20) & $ 13.05$ & $  0.68\pm  0.14$ & $ -1.12\pm  0.10$ & $10.872\pm 0.067$ & $ 0.174\pm 0.010$\\
& [12.60, 12.90) & $ 12.75$ & $  0.43\pm  0.05$ & $ -1.13\pm  0.07$ & $10.746\pm 0.069$ & $ 0.135\pm 0.020$\\
& [12.30, 12.60) & $ 12.45$ & $  0.26\pm  0.04$ & $ -1.21\pm  0.10$ & $10.601\pm 0.085$ & $ 0.127\pm 0.027$\\
& [12.00, 12.30) & $ 12.16$ & $  0.15\pm  0.03$ & $ -1.25\pm  0.14$ & $10.414\pm 0.097$ & $ 0.122\pm 0.030$\\
\enddata

\tablecomments{ Column (1): galaxy type.  Column (2): halo mass range. Column
  (3): average of the logarithm of  the halo mass.  Column (4)-(7): average of
  the best fit free parameters to the four measurements of the CSMFs, as shown
  in  Fig. \ref{fig:fit_CSMF}.  The  errors indicate  the scatter  among these
  four measurements  or the scatter  obtained from the 200  bootstrap samples,
  whichever is larger.}\label{tab:CSMF}
\end{deluxetable*}

In paper  II, we  have measured the  conditional luminosity function  (CLF) of
galaxies in halos (as represented by galaxy groups), $\Phi(L \vert M_h)$. Here
we first  obtain the conditional stellar  mass function (CSMF)  of galaxies in
dark halos.  The CSMF, $\Phi(M_{\ast} \vert M_h)$, which describes the average
number of galaxies as a function of  galaxy stellar mass in a dark matter halo
of a given mass, is  more straightforwardly related to theoretical predictions
of galaxy formation  models than the CLF, because  the conversion from stellar
mass to  luminosity in  theoretical models requires  detailed modeling  of the
stellar population and dust extinction.  The CSMF can be estimated by directly
counting  the  number  of  galaxies   in  groups.   However,  because  of  the
completeness limits discussed in  Section \ref{sec:comp}, we only use galaxies
and  groups  that  are  complete  according  to  Eqs.~(\ref{eq:mstarlim})  and
(\ref{eq:Mh_limit})  to estimate the  CSMF, $\Phi(M_{\ast}  \vert M_h)$,  at a
given  $M_{\ast}$.  In  Fig.~\ref{fig:CSMF} we  show the  resulting  CSMFs for
groups  of  different masses.   The  contributions  of  central and  satellite
galaxies are plotted separately.  For comparison, results obtained using $M_S$
and $M_L$ are shown as symbols and dashed lines, respectively.  The error-bars
shown  in  each panel  correspond  to  1-$\sigma$  scatter obtained  from  200
bootstrap samples of  our group catalogue.  In general,  these two halo masses
give  consistent results,  except that  the  $M_S$-based CSMF  of the  central
galaxies in low  mass halos is more peaked than the  $M_L$-based CSMF (see the
lower right-hand panel).  The general behavior  of the CSMF is similar to that
of the  CLF presented in Paper II.  The general behavior of  the CSMF obtained
here is also qualitatively similar to the prediction of semi-analytical models
(e.g. Zheng et  al. 2005): the CSMF for  small halos has a strong  peak at the
bright end  due to central galaxies.  Quantitatively, however, semi-analytical
models in general over-predict the number of satellite galaxies (Liu et al. in
preparation).

In  Fig.~\ref{fig:CSMF_color} we  show the  CSMFs separately  for  red (dashed
lines with  errorbars) and  blue (dotted lines)  galaxies.  Clearly  there are
more red galaxies than blue galaxies (both centrals and satellites) in massive
halos.  In  the lowest  mass bin probed  here ($12.0  < \log M_h  \leq 12.3$),
however,  there are  roughly  equal numbers  of  red and  blue galaxies.   The
fraction of red galaxies as a function of halo mass found here is very similar
to  that  obtained by  Zandivarez  et al.   (2006)  based  on the  conditional
luminosity   function  of   galaxies   derived  from   an  independent   group
catalog. Note that  the overall shapes of the CSMFs for  red and blue galaxies
are remarkably  similar.  Interestingly, such behavior is  predicted by Skibba
et al.  (2008) who used the color-marked correlation function to constrain the
distribution of galaxies according to their colors.

We model the CSMF using the sum of the CSMFs of central and satellite galaxies
(see Yang et  al. 2003; Cooray 2005; White \etal 2007;  Zheng \etal 2007; Yang
et al. 2008; Cacciato et al. 2008):
\begin{equation}\label{eq:CSMF_fit}
\Phi(M_{\ast}|M_h) = \Phi_{\rm cen}(M_{\ast}|M_h) + \Phi_{\rm sat}(M_{\ast}|M_h)\,.
\end{equation}
Following  Paper II,  we  adopt a  lognormal  model for  the  CSMF of  central
galaxies:
\begin{equation}\label{eq:phi_c}
\Phi_{\rm cen}(M_{\ast}|M_h) = {A\over {\sqrt{2\pi}\sigma_c}} {\rm exp}
\left[- { {(\log M_{\ast}  -\log M_{\ast,c} )^2 } \over 2\sigma_c^2} \right]\,,
\end{equation}
where $A$ is  the number of central galaxies per halo.   Thus, $A\equiv 1$ for
all  galaxies,   $A=f_{\rm  red}(M_h)$  for  red   galaxies,  and  $A=1-f_{\rm
red}(M_h)$ for blue galaxies.  Here  $f_{\rm red}(M_h)$ is the red fraction of
central galaxies in  halos of mass $M_h$.  Note that  $\log M_{\ast,c}$ is, by
definition, the expectation value for  the (10-based) logarithm of the stellar
mass of the central galaxy:
\begin{equation}
\log M_{\ast,c} = \int_0^{\infty} \Phi_{\rm cen}(M_{\ast}|M_h) \log M_{\ast}
{\rm d}\log M_{\ast}\,,
\end{equation}
and  that $\sigma_c=\sigma(\log  M_{\ast})$.   For the  contribution from  the
satellite galaxies we adopt a modified Schechter function:
\begin{equation}\label{eq:phi_s}
\Phi_{\rm sat}(M_{\ast}|M_h) = \phi^*_s \left 
   ( {M_{\ast}\over M_{\ast,s}}\right )^{(\alpha^*_s+1)}
  {\rm exp} \left[- \left ({M_{\ast}\over M_{\ast,s}}\right )^2 \right]\,.
\end{equation}
Note that  this function decreases faster  at the bright end  than a Schechter
function   and  gives   a  better   description  of   the  data.    The  above
parameterization  has   a  total   of  five  free   parameters:  $M_{\ast,c}$,
$\sigma_c$,  $\phi^*_s$, $\alpha^*_s$  and $M_{\ast,s}$.   We find  that $\log
M_{\ast,c} =  \log M_{\ast,s} +0.25$ to  good approximation, which  is what we
adopt throughout.  Consequently,  the number of free parameters  is reduced to
four.  Note, however,  Hansen et al.  (2007) found that  the ratio between the
mean stellar  mass (luminosity) of  the central galaxy and  the characteristic
stellar  mass (luminosity)  of the  satellite  galaxies depends  on halo  mass
especially for  massive halos.   The difference may  arise from the  fact that
they are  using galaxy groups  constructed from the SDSS  photometric redshift
catalogue where the memberships of the groups are not so well constrained.

We  fit   the  above  model   to  all  our   CSMFs.   Results  are   shown  in
Fig~\ref{fig:fit_CSMF}, separately for all  (left panels), red (middle panels)
and blue (right panels) galaxies.   Here comparison is made between samples II
and III.   Note that  Sample II does  not include  any galaxies missed  due to
fiber collisions,  while Sample  III includes all  such galaxies  by assigning
each of them the redshift of its nearest neighbor. Given that we also have two
kinds of  halo masses, $M_L$  and $M_S$, there  are a total of  four different
combinations  for  which   we  have  determined  the  CSMFs.   Each  panel  of
Fig~\ref{fig:fit_CSMF}  shows   the  results  for  all  four   cases.   As  an
illustration  of  how  well the  model  fits  the  data,  the solid  lines  in
Figs.~\ref{fig:CSMF} and ~\ref{fig:CSMF_color} show the corresponding best-fit
models.

The upper row  of Fig.~\ref{fig:fit_CSMF} shows the best  fit normalization of
the  CSMF  for satellite  galaxies,  which  describes  the average  number  of
satellite galaxies with  stellar mass $\sim M_{\ast,s}$ in a  group of a given
halo mass.  As expected, Sample  III gives a higher $\phi^*_s$, especially for
low-mass groups, but only marginally  so.  Comparing $\phi^*_s$ for red (upper
middle  panel)  and blue  (upper  right panel)  galaxies,  one  sees that  the
fraction of red satellites increases with halo mass.  The second row shows the
faint end slopes  of the CSMFs, $\alpha^*_s$.  In massive  halos with $M_h \ga
10^{13}\msunh$,  $\alpha^*_s$  decreases (i.e.,  becomes  more negative)  with
increasing halo mass, both for red  and blue galaxies.  In halos with $M_h \la
10^{13}\msunh$, however, $\alpha^*_s$ is roughly constant at $\sim -1.2$.  The
third row shows that $\log M_{\ast, c}$ increases with halo mass, for both red
and blue centrals. Note that for a given value of $\log M_{\ast, c}$, the halo
mass for  blue galaxies  based on $M_L$  is larger  than that based  on $M_S$,
especially in small halos.  The reason is that for a given stellar mass, bluer
galaxies are  brighter, hence $M_L$ is  higher.  This effect  can impact (e.g.
slightly  enhance) our  results on  the color  dependence of  group clustering
(e.g., Berlind et  al.  2006; Wang et al.  2008a).   These best fit parameters
for satellite  galaxies of different  colors are different from  that obtained
from the  CLF measurements (e.g. Zandivarez  et al. 2006; Hansen  et al. 2007;
Yang  et  al. 2008).  In  particular  the  low-mass end  slope,  $\alpha^*_s$,
obtained  here for  red satellite  galaxies  in small  halos is  significantly
steeper. The only  significant difference between red and  blue galaxies is in
$\phi^*_s$, consistent  with the results  of Skibba (2008). Finally,  the last
row  of Fig  \ref{fig:fit_CSMF}  shows the  width  of the  log-normal CSMF  of
central galaxies.  For the combined sample of red and blue galaxies we find an
average value of $\sigma_c =  \sigma(\log M_{\ast,c}) \sim 0.17$ in halos with
$M_h  \ga 10^{13}\msunh$.   This width  is slightly  larger than  that  in the
luminosity distribution  obtained in Paper II.  A  roughly constant dispersion
in  the lognormal  distribution of  luminosity (or  stellar mass)  for central
galaxies has  already been predicted by  Yang \etal (2003),  Cooray (2006) and
Cacciato  et  al. (2008)  from  the  clustering  and abundances  of  galaxies.
Although Zheng et al.  (2007) found, based  on a HOD model applied to the SDSS
and DEEP2,  that the log-normal width  increases from $\sim  0.13$ for massive
halos with  $M_h \sim 10^{13.5}\msunh$ to  $\sim 0.3$ for low  mass halos with
$M_h \sim 10^{11.5}\msunh$,  the difference between their results  and ours is
only  at 1-$\sigma$  level.  Again,  since our  halo masses  are based  on the
ranking of either $L_{19.5}$ or $M_{\rm stellar}$, we have effectively assumed
a one-to-one relation  between halo mass and these  mass indicators.  This can
give rise  to spurious correlation  between $L_{19.5}$ (or  $M_{\rm stellar}$)
with the stellar  mass of the central galaxies,  especially for low-mass halos
where the stellar  content is dominated by the  central galaxy.  Therefore the
values   of  $\sigma_c$  obtained,   especially  for   halos  with   $M_h  \la
10^{13}\msunh$,  may  be underestimated  and  should  be  considered as  lower
limits.

For reference, Table~\ref{tab:CSMF} lists the {\it average} values of the CSMF
fitting parameters  obtained from the combinations  of Samples II  and III and
group masses $M_L$ and $M_S$.  The error-bars indicate the scatter among these
four samples or the scatter  obtained from 200 bootstrap samples, whichever is
larger (generally  the former).

\section{The group luminosity and stellar-mass functions} 
\label{sec_LF_grp}

\begin{figure*}
  \plotone{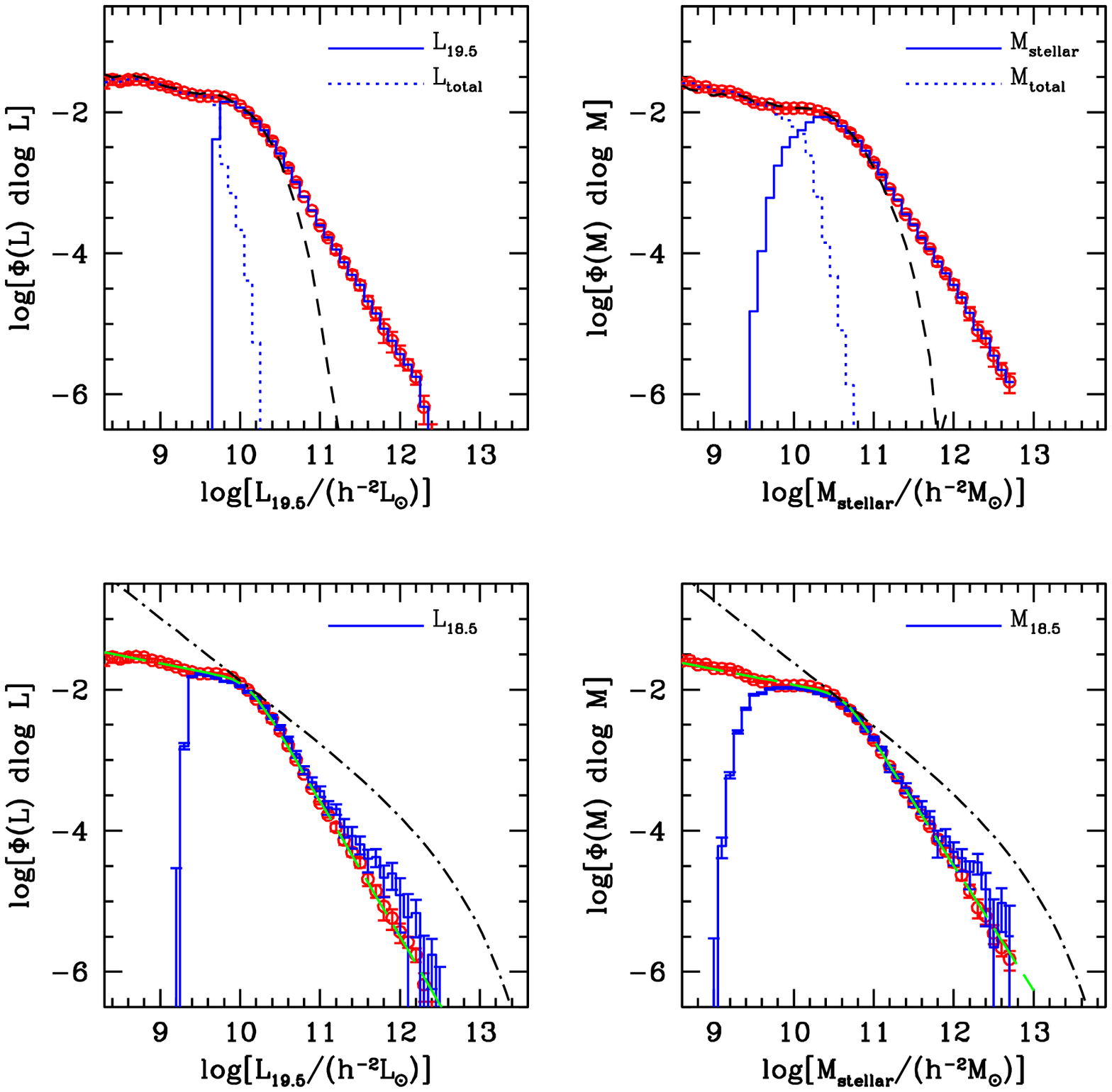}
  \caption{The  group luminosity  functions and  stellar mass  functions.  The
    upper-left  panel: solid and  dotted histograms  are the  group luminosity
    functions  for $L_{19.5}$  and  $L_{\rm total}$,  respectively.  The  open
    circles with error-bars are the  sum of the two contributions.  The dashed
    line  is  the luminosity  function  for  central  galaxies shown  in  Fig.
    \ref{fig:gax_LF_MF}.  The  upper-right panel: solid  and dotted histograms
    are the  group stellar  mass functions for  $M_{\rm stellar}$  and $M_{\rm
      total}$, respectively.  The open circles  with error-bars are the sum of
    the two  contributions. The dashed line  is the stellar  mass function for
    central galaxies shown in Fig.~\ref{fig:gax_LF_MF}.  The lower-left panel:
    the open circles  with error-bars are the same as  those in the upper-left
    panel and the long dashed line  shows the best fitting results.  The solid
    histogram is the group luminosity function for $L_{18.5}$. The lower-right
    panel:  the open  circles with  error-bars are  the same  as those  in the
    upper-right  panel  and  the  long  dashed line  shows  the  best  fitting
    results.  The solid  histogram  is  the group  stellar  mass function  for
    $M_{18.5}$.  See text for  the definitions of $L_{19.5}$, $L_{\rm total}$,
    $L_{18.5}$,  $M_{\rm  stellar}$,   $M_{\rm  total}$  and  $M_{18.5}$.  For
    comparison,  in the  lower two  panels, we  show as  dot-dashed  lines the
    scaled halo mass function (Warren  et al.  2006) for the same $\Lambda$CDM
    `concordance' cosmology. }
\label{fig:grp_LF_MF}
\end{figure*}

\begin{deluxetable*}{ccccccc}
  \tabletypesize{\scriptsize} 
  \tablecaption{The group luminosity functions and stellar mass functions} 
  \tablewidth{0pt} 
  \tablehead{ $\log L$ ($\log M_{\ast}$) & $\Phi(L_{\rm total})$ & $\Phi(L_{19.5})$ &
  $\Phi(L_{18.5})$ & $\Phi(M_{\rm total})$ & $\Phi(M_{\rm stellar})$ & $\Phi(M_{18.5})$ \\ 
  \cline{1-7}\\ (1) & (2) & (3) & (4) & (5) & (6) & (7)} 

\startdata
  8.1  &  3.2253 $\pm$ 0.5322  &  0.0000 $\pm$ 0.0000  &  0.0000 $\pm$ 0.0000  &  0.0000 $\pm$ 0.0000  &  0.0000 $\pm$ 0.0000  &  0.0000 $\pm$ 0.0000 \\
  8.2  &  3.4201 $\pm$ 0.6087  &  0.0000 $\pm$ 0.0000  &  0.0000 $\pm$ 0.0000  &  0.0000 $\pm$ 0.0000  &  0.0000 $\pm$ 0.0000  &  0.0000 $\pm$ 0.0000 \\
  8.3  &  2.6641 $\pm$ 0.5050  &  0.0000 $\pm$ 0.0000  &  0.0000 $\pm$ 0.0000  &  0.0000 $\pm$ 0.0000  &  0.0000 $\pm$ 0.0000  &  0.0000 $\pm$ 0.0000 \\
  8.4  &  2.8979 $\pm$ 0.4862  &  0.0000 $\pm$ 0.0000  &  0.0000 $\pm$ 0.0000  &  2.3283 $\pm$ 0.4981  &  0.0000 $\pm$ 0.0000  &  0.0000 $\pm$ 0.0000 \\
  8.5  &  2.7062 $\pm$ 0.2931  &  0.0000 $\pm$ 0.0000  &  0.0000 $\pm$ 0.0000  &  2.2944 $\pm$ 0.3343  &  0.0000 $\pm$ 0.0000  &  0.0000 $\pm$ 0.0000 \\
  8.6  &  2.8626 $\pm$ 0.3414  &  0.0000 $\pm$ 0.0000  &  0.0000 $\pm$ 0.0000  &  2.5619 $\pm$ 0.1969  &  0.0000 $\pm$ 0.0000  &  0.0000 $\pm$ 0.0000 \\
  8.7  &  2.9337 $\pm$ 0.1897  &  0.0000 $\pm$ 0.0000  &  0.0000 $\pm$ 0.0000  &  2.5771 $\pm$ 0.1779  &  0.0000 $\pm$ 0.0000  &  0.0000 $\pm$ 0.0000 \\
  8.8  &  2.8651 $\pm$ 0.1770  &  0.0000 $\pm$ 0.0000  &  0.0000 $\pm$ 0.0000  &  2.2581 $\pm$ 0.1293  &  0.0000 $\pm$ 0.0000  &  0.0000 $\pm$ 0.0000 \\
  8.9  &  2.5868 $\pm$ 0.1677  &  0.0000 $\pm$ 0.0000  &  0.0000 $\pm$ 0.0000  &  2.2786 $\pm$ 0.1022  &  0.0000 $\pm$ 0.0000  &  0.0000 $\pm$ 0.0000 \\
  9.0  &  2.4589 $\pm$ 0.2062  &  0.0000 $\pm$ 0.0000  &  0.0000 $\pm$ 0.0000  &  2.0151 $\pm$ 0.0913  &  0.0000 $\pm$ 0.0000  &  0.0000 $\pm$ 0.0003 \\
  9.1  &  2.2798 $\pm$ 0.1883  &  0.0000 $\pm$ 0.0000  &  0.0000 $\pm$ 0.0000  &  1.9811 $\pm$ 0.1045  &  0.0000 $\pm$ 0.0000  &  0.0060 $\pm$ 0.0020 \\
  9.2  &  2.1088 $\pm$ 0.1361  &  0.0000 $\pm$ 0.0000  &  0.0000 $\pm$ 0.0030  &  1.9434 $\pm$ 0.1014  &  0.0000 $\pm$ 0.0000  &  0.0613 $\pm$ 0.0065 \\
  9.3  &  1.8898 $\pm$ 0.0820  &  0.0000 $\pm$ 0.0000  &  0.1574 $\pm$ 0.0163  &  1.7766 $\pm$ 0.1248  &  0.0000 $\pm$ 0.0000  &  0.2513 $\pm$ 0.0117 \\
  9.4  &  1.7805 $\pm$ 0.1027  &  0.0000 $\pm$ 0.0000  &  1.5209 $\pm$ 0.0291  &  1.5740 $\pm$ 0.1087  &  0.0000 $\pm$ 0.0004  &  0.5356 $\pm$ 0.0192 \\
  9.5  &  1.6742 $\pm$ 0.0629  &  0.0000 $\pm$ 0.0000  &  1.6765 $\pm$ 0.0450  &  1.3785 $\pm$ 0.0983  &  0.0015 $\pm$ 0.0028  &  0.8491 $\pm$ 0.0229 \\
  9.6  &  1.6838 $\pm$ 0.2660  &  0.0000 $\pm$ 0.1849  &  1.5925 $\pm$ 0.0407  &  1.2934 $\pm$ 0.0800  &  0.0107 $\pm$ 0.0160  &  0.8836 $\pm$ 0.0238 \\
  9.7  &  1.2575 $\pm$ 0.2790  &  0.4140 $\pm$ 0.2468  &  1.4683 $\pm$ 0.0367  &  1.2237 $\pm$ 0.0737  &  0.0610 $\pm$ 0.0349  &  1.0167 $\pm$ 0.0220 \\
  9.8  &  0.1845 $\pm$ 0.2831  &  1.4048 $\pm$ 0.2795  &  1.3138 $\pm$ 0.0266  &  0.9211 $\pm$ 0.0728  &  0.1742 $\pm$ 0.0574  &  1.0525 $\pm$ 0.0230 \\
  9.9  &  0.0713 $\pm$ 0.0668  &  1.3972 $\pm$ 0.1135  &  1.2828 $\pm$ 0.0312  &  0.7662 $\pm$ 0.0597  &  0.3168 $\pm$ 0.0601  &  1.0576 $\pm$ 0.0227 \\
 10.0  &  0.0213 $\pm$ 0.0209  &  1.1648 $\pm$ 0.0569  &  1.1205 $\pm$ 0.0360  &  0.6207 $\pm$ 0.0561  &  0.4414 $\pm$ 0.0561  &  1.0479 $\pm$ 0.0220 \\
 10.1  &  0.0041 $\pm$ 0.0025  &  0.9810 $\pm$ 0.0490  &  0.9278 $\pm$ 0.0323  &  0.4901 $\pm$ 0.0900  &  0.5571 $\pm$ 0.0562  &  1.0148 $\pm$ 0.0185 \\
 10.2  &  0.0005 $\pm$ 0.0014  &  0.7398 $\pm$ 0.0543  &  0.7652 $\pm$ 0.0244  &  0.2426 $\pm$ 0.0767  &  0.7160 $\pm$ 0.0428  &  0.9849 $\pm$ 0.0218 \\
 10.3  &  0.0000 $\pm$ 0.0001  &  0.5534 $\pm$ 0.0597  &  0.5713 $\pm$ 0.0254  &  0.0630 $\pm$ 0.0526  &  0.8471 $\pm$ 0.0456  &  0.8950 $\pm$ 0.0262 \\
 10.4  &  0.0000 $\pm$ 0.0000  &  0.3862 $\pm$ 0.0426  &  0.4079 $\pm$ 0.0239  &  0.0140 $\pm$ 0.0097  &  0.8444 $\pm$ 0.0435  &  0.8355 $\pm$ 0.0267 \\
 10.5  &  0.0000 $\pm$ 0.0000  &  0.2608 $\pm$ 0.0242  &  0.2914 $\pm$ 0.0242  &  0.0048 $\pm$ 0.0007  &  0.7618 $\pm$ 0.0621  &  0.7352 $\pm$ 0.0218 \\
 10.6  &  0.0000 $\pm$ 0.0000  &  0.1620 $\pm$ 0.0140  &  0.1987 $\pm$ 0.0145  &  0.0008 $\pm$ 0.0005  &  0.6367 $\pm$ 0.0635  &  0.6321 $\pm$ 0.0159 \\
 10.7  &  0.0000 $\pm$ 0.0000  &  0.1020 $\pm$ 0.0069  &  0.1215 $\pm$ 0.0131  &  0.0001 $\pm$ 0.0003  &  0.5056 $\pm$ 0.0562  &  0.5197 $\pm$ 0.0201 \\
 10.8  &  0.0000 $\pm$ 0.0000  &  0.0635 $\pm$ 0.0012  &  0.0731 $\pm$ 0.0094  &  0.0000 $\pm$ 0.0001  &  0.3893 $\pm$ 0.0433  &  0.3832 $\pm$ 0.0225 \\
 10.9  &  0.0000 $\pm$ 0.0000  &  0.0400 $\pm$ 0.0008  &  0.0488 $\pm$ 0.0078  &  0.0000 $\pm$ 0.0000  &  0.2816 $\pm$ 0.0307  &  0.2774 $\pm$ 0.0203 \\
 11.0  &  0.0000 $\pm$ 0.0000  &  0.0249 $\pm$ 0.0011  &  0.0363 $\pm$ 0.0062  &  0.0000 $\pm$ 0.0000  &  0.1918 $\pm$ 0.0183  &  0.2025 $\pm$ 0.0127 \\
 11.1  &  0.0000 $\pm$ 0.0000  &  0.0167 $\pm$ 0.0015  &  0.0215 $\pm$ 0.0048  &  0.0000 $\pm$ 0.0000  &  0.1295 $\pm$ 0.0106  &  0.1460 $\pm$ 0.0079 \\
 11.2  &  0.0000 $\pm$ 0.0000  &  0.0112 $\pm$ 0.0014  &  0.0202 $\pm$ 0.0033  &  0.0000 $\pm$ 0.0000  &  0.0817 $\pm$ 0.0048  &  0.0751 $\pm$ 0.0079 \\
 11.3  &  0.0000 $\pm$ 0.0000  &  0.0074 $\pm$ 0.0011  &  0.0113 $\pm$ 0.0033  &  0.0000 $\pm$ 0.0000  &  0.0566 $\pm$ 0.0019  &  0.0568 $\pm$ 0.0066 \\
 11.4  &  0.0000 $\pm$ 0.0000  &  0.0050 $\pm$ 0.0007  &  0.0084 $\pm$ 0.0028  &  0.0000 $\pm$ 0.0000  &  0.0358 $\pm$ 0.0016  &  0.0404 $\pm$ 0.0067 \\
 11.5  &  0.0000 $\pm$ 0.0000  &  0.0035 $\pm$ 0.0006  &  0.0064 $\pm$ 0.0018  &  0.0000 $\pm$ 0.0000  &  0.0253 $\pm$ 0.0007  &  0.0263 $\pm$ 0.0044 \\
 11.6  &  0.0000 $\pm$ 0.0000  &  0.0021 $\pm$ 0.0004  &  0.0040 $\pm$ 0.0014  &  0.0000 $\pm$ 0.0000  &  0.0166 $\pm$ 0.0009  &  0.0218 $\pm$ 0.0046 \\
 11.7  &  0.0000 $\pm$ 0.0000  &  0.0014 $\pm$ 0.0003  &  0.0043 $\pm$ 0.0012  &  0.0000 $\pm$ 0.0000  &  0.0115 $\pm$ 0.0008  &  0.0158 $\pm$ 0.0032 \\
 11.8  &  0.0000 $\pm$ 0.0000  &  0.0008 $\pm$ 0.0003  &  0.0022 $\pm$ 0.0010  &  0.0000 $\pm$ 0.0000  &  0.0076 $\pm$ 0.0010  &  0.0071 $\pm$ 0.0030 \\
 11.9  &  0.0000 $\pm$ 0.0000  &  0.0006 $\pm$ 0.0002  &  0.0025 $\pm$ 0.0010  &  0.0000 $\pm$ 0.0000  &  0.0052 $\pm$ 0.0005  &  0.0083 $\pm$ 0.0022 \\
 12.0  &  0.0000 $\pm$ 0.0000  &  0.0004 $\pm$ 0.0001  &  0.0012 $\pm$ 0.0009  &  0.0000 $\pm$ 0.0000  &  0.0036 $\pm$ 0.0005  &  0.0050 $\pm$ 0.0018 \\
 12.1  &  0.0000 $\pm$ 0.0000  &  0.0003 $\pm$ 0.0000  &  0.0006 $\pm$ 0.0007  &  0.0000 $\pm$ 0.0000  &  0.0023 $\pm$ 0.0003  &  0.0041 $\pm$ 0.0014 \\
 12.2  &  0.0000 $\pm$ 0.0000  &  0.0002 $\pm$ 0.0000  &  0.0007 $\pm$ 0.0004  &  0.0000 $\pm$ 0.0000  &  0.0014 $\pm$ 0.0003  &  0.0026 $\pm$ 0.0011 \\
 12.3  &  0.0000 $\pm$ 0.0000  &  0.0001 $\pm$ 0.0000  &  0.0000 $\pm$ 0.0003  &  0.0000 $\pm$ 0.0000  &  0.0008 $\pm$ 0.0002  &  0.0035 $\pm$ 0.0012 \\
 12.4  &  0.0000 $\pm$ 0.0000  &  0.0000 $\pm$ 0.0000  &  0.0002 $\pm$ 0.0001  &  0.0000 $\pm$ 0.0000  &  0.0006 $\pm$ 0.0002  &  0.0015 $\pm$ 0.0010 \\
 12.5  &  0.0000 $\pm$ 0.0000  &  0.0000 $\pm$ 0.0000  &  0.0000 $\pm$ 0.0001  &  0.0000 $\pm$ 0.0000  &  0.0004 $\pm$ 0.0001  &  0.0002 $\pm$ 0.0008 \\
 12.6  &  0.0000 $\pm$ 0.0000  &  0.0000 $\pm$ 0.0000  &  0.0000 $\pm$ 0.0000  &  0.0000 $\pm$ 0.0000  &  0.0002 $\pm$ 0.0001  &  0.0010 $\pm$ 0.0006 \\
 12.7  &  0.0000 $\pm$ 0.0000  &  0.0000 $\pm$ 0.0000  &  0.0000 $\pm$ 0.0000  &  0.0000 $\pm$ 0.0000  &  0.0002 $\pm$ 0.0000  &  0.0003 $\pm$ 0.0006 \\
 \enddata
 
 \tablecomments{Column  (1):  the  median   of  the  logarithm  of  the  group
   luminosity or stellar mass with  bin width $\Delta \log L=0.05$ (or $\Delta
   \log M_{\ast}=0.05$  ).  Column (2): the average  group luminosity function
   for $L_{\rm total}$,  where $L_{\rm total}$ is the  total luminosity of all
   group members in  which the central galaxy has luminosity  $\rmag > -19.5$. 
   Column  (3): the average  group luminosity  function for  $L_{19.5}$, where
   $L_{19.5}$ is the  luminosity of all group members with  $\rmag \le -19.5$. 
   Column  (4) the  average group  luminosity function  for  $L_{18.5}$, where
   $L_{18.5}$ is  the total  luminosity of all  group members with  $\rmag \le
   -18.5$. Columns  (5)-(7): Similar  to Columns (2)-(4)  but for  the average
   stellar mass functions.  In this  table, the averages are obtained from the
   two measurements of the luminosity functions or stellar mass functions from
   samples II  and III,  respectively. The error  indicates the  scatter among
   these two measurements or the  scatter obtained from 200 bootstrap samples,
   whichever is  larger.  Note  that all the  group luminosity  (stellar mass)
   functions listed in this table  are calculated in units of $10^{-2} h^3{\rm
     Mpc}^{-3} {\rm  d} \log  L$ (or $10^{-2}  h^3{\rm Mpc}^{-3} {\rm  d} \log
   M_{\ast}$), where $\log$ is the 10 based logarithm.   
}\label{tab:grp_LF_MF}
\end{deluxetable*}

In  this  section,  we  present  our  results  on  the  group  luminosity  and
stellar-mass functions.   These two functions depend the  total luminosity and
stellar mass in  galaxy groups, and are arguably  better suited for comparison
with  model predictions, as  the details  about how  the total  luminosity and
total  stellar mass  are partitioned  into individual  member galaxies  is not
important here.   We measure the  group luminosity and  stellar-mass functions
for both samples II and III.   As mentioned above, these two samples represent
two extremes,  as far as  fiber-collision effects are concerned.   The results
presented in  the following  are based on  the average  of samples II  and III
(evenly weighted), and the error-bars are obtained from the difference between
the  two samples,  or  from 200  bootstrap  samples for  samples  II and  III,
whichever (typically the difference between the two samples) is larger.

As  discussed in  Section  \ref{sec:comp}, the  group  catalogue suffers  from
incompleteness, in that groups within certain luminosity ($L_{19.5}$), stellar
mass  ($M_{\rm stellar}$)  and halo  mass bins  are only  complete  to certain
redshifts. Using the same procedure as described in Section \ref{sec:comp}, we
estimate,  for  each group  luminosity  $L_{19.5}$  (or  stellar mass  $M_{\rm
stellar}$),   the   maximum    redshift,   $z_{\rm   max}$,   that   satisfies
Eq. (\ref{eq:n_z}).   The resulting  $n(z_{\rm max})$ as  a function  of group
luminosity  $L_{19.5}$ or  stellar  mass $M_{\rm  stellar}$  gives, the  group
luminosity  function  $\Phi(L_{19.5})$  or  the group  stellar  mass  function
$\Phi(M_{\rm  stellar})$, respectively.  The  results are  shown as  the solid
histograms in the upper panels of Fig.~\ref{fig:grp_LF_MF}.

Note  that the  group characteristic  luminosity $L_{19.5}$  and  stellar mass
$M_{\rm stellar}$ are  defined to be the total luminosity  and stellar mass of
member galaxies with  $\rmag \le -19.5$.  There are  certainly groups in which
all member  galaxies have  $\rmag > -19.5$,  so that $L_{19.5}=0$  and $M_{\rm
stellar}=0$.  For {\it these} groups we measure their total luminosity $L_{\rm
total}$ and  total stellar mass $M_{\rm  total}$ based on  all member galaxies
that are  observed.  We measure the luminosity  function $\Phi(L_{\rm total})$
and stellar  mass function  $\Phi(M_{\rm total})$ for  these groups  using the
same  method as  for $\Phi(L_{19.5})$  and $\Phi(M_{\rm  stellar})$,  i.e., by
measuring the $n(z_{\rm max})$.  The results  are shown in the upper panels of
Fig \ref{fig:grp_LF_MF} as the dotted histograms. For comparison, we also show
as the dashed lines the galaxy luminosity function (stellar mass function) for
central galaxies that we  obtained in Section \ref{sec_LF_gax}.  The agreement
between  the group  luminosity  function $\Phi(L_{\rm  total})$ (stellar  mass
function  $\Phi(M_{\rm total})$)  and the  central galaxy  luminosity function
(stellar mass function) at the faint (low-mass) end is excellent.  This simply
reflects  that the  total luminosity  $L_{\rm total}$  and total  stellar mass
$M_{\rm  total}$ of  small  groups  are dominated  by  their central  galaxies
(e.g. Lin \&  Mohr 2004; Hansen et  al. 2007; Yang et al.  2008b), and implies
that missing faint satellites in low  mass groups has no significant impact on
our estimates of the group luminosity and stellar-mass functions, unless there
is a sharp upturn in the  conditional luminosity (or stellar mass) function of
satellite  galaxies,  as  indicated  in  some observations  (e.g.  Popesso  et
al. 2006). Thus the overall  group luminosity function (stellar mass function)
from small  to large  groups can  be obtained by  sum up  $\Phi(L_{19.5})$ and
$\Phi(L_{\rm  total})$, ($\Phi(M_{\rm  stellar})$ and  $\Phi(M_{\rm total})$).
The results are shown as the open circles in the upper panels.

As a further test of the  reliability of our results at the low-luminosity and
low-mass end,  we use another  characteristic group luminosity  $L_{18.5}$ and
another characteristic stellar mass $M_{18.5}$ to represent the luminosity and
stellar mass  of groups.   Here $L_{18.5}$ ($M_{18.5}$)  is defined to  be the
total luminosity (stellar  mass) of all group members with  $\rmag \le -18.5$. 
Because of the survey apparent magnitude limit, a galaxy fainter than $\rmag =
-18.5$ is not  observed if its redshift is larger than  $\sim 0.06$.  To avoid
such incompleteness,  $L_{18.5}$ and $M_{18.5}$  are only measured  for groups
with  redshift  $z\le 0.06$.   Our  method  to  estimate $\Phi(L_{18.5})$  and
$\Phi(M_{18.5})$  is   the  same  as   that  used  for   $\Phi(L_{19.5})$  and
$\Phi(M_{\rm stellar})$, except that  $z_{\rm max}$ starts from $0.06$ instead
of $z=0.2$.   The results for $\Phi(L_{18.5})$ and  $\Phi(M_{18.5})$ are shown
in the lower  two panels of Fig.~\ref{fig:grp_LF_MF} as  the solid histograms. 
For  comparison,  we  also  show  $L_{19.5}$  and  $M_{\rm  stellar}$  in  the
corresponding  panels as  open circles.   Clearly, there  is a  good agreement
between  $\Phi(L_{18.5})$ and  $\Phi(L_{19.5})$, and  between $\Phi(M_{18.5})$
and  $\Phi(M_{\rm  stellar})$ for  small  groups,  suggesting  again that  our
results for low-mass  groups are reliable.  For massive  groups, there is some
difference between the two estimates. Unfortunately, the number of such groups
is quite  small in the volume  corresponding to $z=0.06$, and  the results for
$\Phi(L_{18.5})$ and $\Phi(M_{18.5})$ are quite  uncertain at the massive end. 
For reference,  the data for the  group luminosity functions  and stellar mass
functions     shown    in     Fig.~\ref{fig:grp_LF_MF}    are     listed    in
Table~\ref{tab:grp_LF_MF}.

Since both the luminosity function and stellar mass function for groups show
double power-law behavior, we use the following forms to fit the data:
\begin{equation}\label{eq:LF_fit}
\Phi(L_{19.5}) = \Phi_0  \frac {( L_{19.5}/L_0)^\alpha} 
{(x_0+(L_{19.5}/L_0)^4)^\beta}  \,,
\end{equation}
and 
\begin{equation}\label{eq:MF_fit}
\Phi(M_{\rm stellar}) = \Phi_0  \frac {(M_{\rm  stellar}/M_0)^\alpha }
{(x_0+(M_{\rm stellar}/M_0)^4)^\beta }  \,.
\end{equation}
The  best   fit  results  are   shown  in  the   lower  two  panels   of  Fig.
\ref{fig:grp_LF_MF}  as the  long-dashed  lines.  The  corresponding best  fit
parameters  are $[\Phi_0,  \log L_0,  x_0, \alpha,  \beta]$  =[0.00580, 10.30,
0.1786,  -0.2226,  0.4236]  and  $[\Phi_0,  \log  M_0,  x_0,  \alpha,  \beta]$
=[0.00731,  10.67,   0.7243,  -0.2229,  0.3874  ],  for   the  luminosity  and
stellar-mass  functions,  respectively. Note  that  all  the group  luminosity
functions  and stellar  mass functions  are  calculated in  terms of  $h^3{\rm
Mpc}^{-3}  {\rm d} \log  L$ and  $h^3{\rm Mpc}^{-3}  {\rm d}  \log M_{\ast}$),
where $\log$ is the 10 based logarithm. 

It is interesting  to compare the group luminosity  and stellar-mass functions
with the  halo mass function predicted  by the current CDM  model of structure
formation.  In  the lower-left panel  of Fig~\ref{fig:grp_LF_MF}, we  plot, as
the dot-dashed line,  the halo mass function which is  scaled with a (minimum)
constant mass-to-light ratio, $M_h/L_{19.5}=47 h\Msun/\Lsun$, so that the halo
mass function  touches the  group luminosity function  at $\log  L_{19.5} \sim
10.2$.   Here  the  halo  mass  function  is estimated  using  the  Warren  et
al. (2006)  model and adopting  the $\Lambda$CDM `concordance'  cosmology with
parameters listed  at the  end of  Section 1.  Clearly,  the scaled  halo mass
function  has  a very  different  shape  from  the observed  group  luminosity
function, indicating that the mass-to-light ratio must depend strongly on halo
mass.   The mass-to-light  ratio reaches  its minimum  in groups  (halos) with
$\log L_{19.5}\sim 10.2$, and is larger for both smaller and bigger halos.  In
the lower-right panel of Fig.~\ref{fig:grp_LF_MF} we compare the group stellar
mass  function  (open circles)  with  the halo  mass  function  scaled with  a
constant  halo-to-stellar mass  ratio  (dot-dashed line),  $M_h/M_{stellar}=25
h$. Here the halo-to-stellar mass  ratio reaches its minimum in groups (halos)
with  $\log M_{\rm  stellar} \sim  10.6$, corresponding  to halos  with masses
$M_h\sim 11.9$ (almost exactly the same halo mass that the group mass-to-light
ratio reaches its  minimum), suggesting that star formation  efficiency is the
highest in  such halos.  These results  are in excellent  agreement with those
obtained  from the clustering  properties of  galaxies using  the CLF  and HOD
formalism (Yang et  al. 2003; Tinker et  al 2005; van den Bosch  et al.  2007;
Cacciato et  al. 2008), and  at the relative  massive end with  those obtained
from weak lensing observations (e.g. Sheldon et al. 2007).
\begin{figure*} \plotone{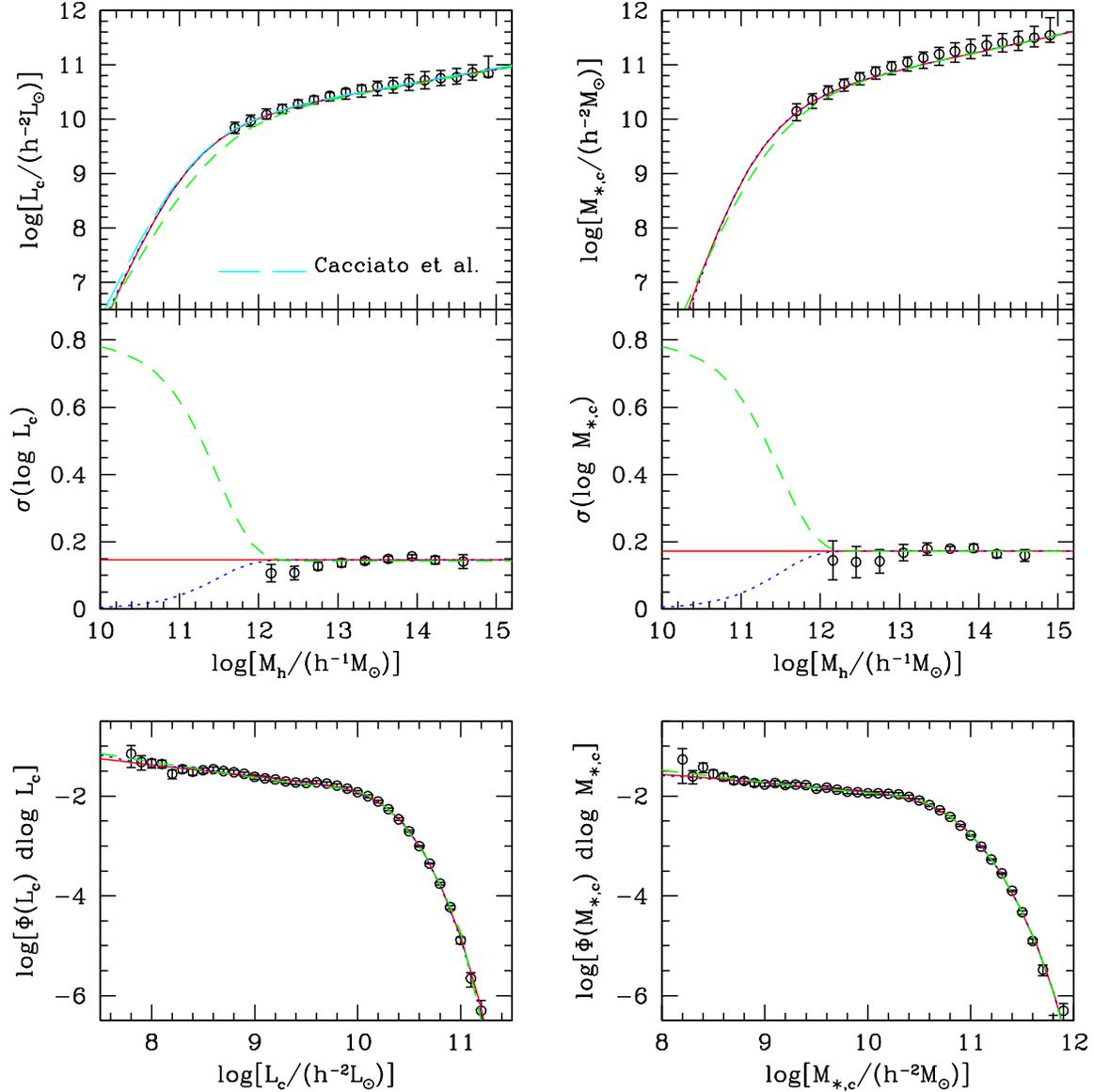}
  \caption{The properties  of central  galaxies: shown in  the left  and right
    panels are  results for  the luminosity $L_c$  and stellar  mass $M_{\ast,
      c}$,  respectively. The  open  circles in  each  panel are  the data  we
    extracted from the  SDSS DR4 group catalogue, while  the solid, dotted and
    dashed  lines  are  the  best  fitting results  (see  text  for  details).
    Upper-left  panel: the $L_c  - M_h$  relation, where  the data  points are
    obtained from the  left panel of Fig. 6 in Paper  II. The long-dashed line
    in  this panel  illustrates  the $L_c  -  M_h$ relation  predicted by  
    Cacciato et al. (2008).   Upper-right  panel:  the  $M_{\ast,  c}  -  M_h$
    relation, where the data points are  obtained from the right panel of Fig.
    6 in Paper  II. Middle-left panel: the $\sigma  (\log L_c) -M_h$ relation,
    where the data points are obtained  from the lower-left panel of Fig. 4 in
    paper  II.   Middle-right  panel:  the  $\sigma  (\log  M_{\ast,c})  -M_h$
    relation, where the data points  are obtained from the lower-left panel of
    Fig.   \ref{fig:fit_CSMF}.  Lower-left panel:  the luminosity  function of
    the central galaxies, $\Phi (L_c)$, where  the data points are the same as
    those  shown  in  the   middle-left  panel  of  Fig.  \ref{fig:gax_LF_MF}.
    Lower-right panel:  the stellar mass  function of central  galaxies, $\Phi
    (M_{\ast,c})$, where  the data points are  the same as those  shown in the
    middle-right panel of Fig. \ref{fig:gax_LF_MF}. }
\label{fig:fit}
\end{figure*}

\section{The central galaxies in small halos}
\label{sec_small}

In this section we discuss the  implication of the observed faint end slope of
the central galaxy luminosity and  stellar mass functions for the relationship
between galaxies and dark matter halos.   In Yang et al. (2005b) and Paper II,
we  have  measured  the  CLFs   for  relatively  massive  halos  with  $M_h\ga
10^{12}\msunh$ directly  from group samples.  Unfortunately,  galaxies in less
massive halos are  not well studied, partly because the  halo masses for small
groups  are difficult  to  estimate.  Here,  with  the help  of the  $L_c-M_h$
($M_{\ast, c} - M_h$) relations  that we obtained for relatively massive halos
and  using the  luminosity functions  and stellar  mass functions  for central
galaxies, we can  probe the halo properties of low-mass  central galaxies in a
statistical   way.     As   discussed   in    Sections   \ref{sec_CSMFs}   and
\ref{sec_LF_grp},  a small  halo  is  usually dominated  by  a single  central
galaxy.   The $L_c-M_h$  and $M_{\ast,  c} -  M_h$ relations  for  groups with
masses down to $\sim 10^{11.8}\msunh$  can be obtained directly from the group
catalogue, and are  shown as open circles with error-bars  in the upper panels
of  Fig.  \ref{fig:fit}.   For these  relatively massive  groups, we  can also
estimate  $\sigma(\log L_c)$ and  $\sigma(\log M_{\ast,  c})$ as  functions of
halo mass.  These are shown in  the middle panels of Fig. \ref{fig:fit} as the
open  circles  with error-bars.   Finally,  the  luminosity  and stellar  mass
functions  for   central  galaxies,  which  have  been   obtained  in  Section
\ref{sec_LF_gax}, are repeated  in the lower panels of  Fig.  \ref{fig:fit} as
open circles with error-bars.

Following  paper II,  we model  the  mean $L_c-M_h$  and $M_{\ast,  c} -  M_h$
relations using the following functional forms:
\begin{equation}\label{eq:Lc_fit}
L_c = L_0 \frac { (M_h/M_1)^{\alpha +\beta} }{(1+M_h/M_1)^\beta } \,,
\end{equation}
and
\begin{equation}\label{eq:Mc_fit}
M_{\ast,c} = M_0 \frac { (M_h/M_1)^{\alpha +\beta} }{(1+M_h/M_1)^\beta } \,,
\end{equation}
where   $M_1$  is   a  characteristic   halo   mass  so   that  $L_c   \propto
M_h^{\alpha+\beta}$  ($M_{\ast,c} \propto  M_h^{\alpha+\beta}$)  for $M_h  \ll
M_1$ and  $L_c \propto  M_h^{\alpha}$ ($M_{\ast,c} \propto  M_h^{\alpha}$) for
$M_h \gg M_1$.  Note that the  parameters $M_1$, $\alpha$ and $\beta$ may have
different values in  the two relations. Clearly, for  bright (massive) central
galaxies, these relations are well constrained by the data points shown in the
upper  panels   of  Fig.    \ref{fig:fit}.   Unfortunately,  no   such  direct
constraints  are available  for  the faint  (low-mass)  centrals. However,  an
indirect  constraint  comes from  the  observed  luminosity  and stellar  mass
functions of central galaxies.  In general, we can write
\begin{equation}\label{eq:model_phiLc}  \Phi(L_c) =  \int_0^{\infty} \Phi_{\rm
cen}(L|M_h) n(M_h) {\rm d}M_h\,,
\end{equation}
and 
\begin{equation}\label{eq:model_phiMc}
\Phi(M_{\ast,c}) = \int_0^{\infty} \Phi_{\rm cen}(M_{\ast}|M_h) n(M_h)
{\rm d} M_h \,,
\end{equation}
where  $n(M)$  is the  mass  function of  dark  matter  halos, and  $\Phi_{\rm
  cen}(L|M_h)$  and  $\Phi_{\rm cen}(M_\ast|M_h)$  are  the  CLF  and CSMF  of
central galaxies,  respectively. If we model both  $\Phi_{\rm cen}(L|M_h)$ and
$\Phi_{\rm  cen}(M_\ast|M_h)$  with  a  lognormal form,  they  are  completely
determined by the mean  relations, (\ref{eq:Lc_fit}) and (\ref{eq:Mc_fit}) and
the corresponding dispersions, $\sigma (\log L_c)$ and $\sigma (\log M_{\ast ,
  c})$. As  a simple model we  first assume these dispersions  to be constant:
$\sigma  (\log L_c)  =\sigma_0$ and  $\sigma  (\log M_{\ast  , c})  =\sigma_0$
(again $\sigma_0$ may have different values in the two cases).  Thus, for each
case, we have five free parameters, $L_0$ (or $M_0$), $M_1$, $\alpha$, $\beta$
and $\sigma_0$.

Using  the halo  mass  function predicted  by  the $\Lambda$CDM  `concordance'
cosmology, we fit  the models described above to  the observational data shown
in  Fig.  \ref{fig:fit}.   The  fitting  is performed  with  a standard  least
$\chi^2$  algorithm.  For  the  luminosity  of the  central  galaxies, we  use
$\chi^2  = \chi^2(L_c)  +  \chi^2(\sigma) +  \chi^2(\Phi(L_c))$,  and for  the
stellar mass  of the central galaxies,  we use $\chi^2  = \chi^2(M_{\ast,c}) +
\chi^2(\sigma)    +     \chi^2(\Phi(M_{\ast,c}))$.     Here,    $\chi^2(L_c)$,
$\chi^2(\sigma)$    and    $\chi^2(\Phi(L_c))$    (or    $\chi^2(M_{\ast,c})$,
$\chi^2(\sigma)$ and  $\chi^2(\Phi(M_{\ast,c}))$) are calculated  according to
the deviations of  the model predictions from the  observational data shown in
the upper-, middle-, and lower-left  (or right) panels of Fig.  \ref{fig:fit},
respectively. Based on all the data  points shown in the left hand side panels
of  Fig. \ref{fig:fit},  we  obtain $\log  L_0=  9.9078$, $\log  M_1=11.0096$,
$\alpha=0.2566$, $\beta  =3.4037$, $\sigma_0=0.1462$.  The  resulting best fit
is shown in each panel as the solid lines.  The agreement between the best fit
model and the  data is remarkably good.  The  best fitting parameters indicate
that $L_c  \propto M_h^{3.7}$ for  $\log M_h \ll  11.0 $, suggesting  that the
star  formation  efficiency  decreases  dramatically in  small  halos.   These
results,  especially the  slopes of  the $L_c$  - $M_h$  relation ($\alpha\sim
0.25$ at the  massive end, and $\alpha+\beta>>1$ at the  low-mass end), are in
good  agreement with  previous results  (e.g.,  Vale \&  Ostriker 2004,  2006;
Cooray 2005;  Yang \etal 2003;  Yang et al.  2005c; van den Bosch  \etal 2007;
Hansen  et al.  2007;  Zheng  et al.  2007;  Popesso et  al.  2007; Brough  et
al. 2008;  Conroy \& Wechsler  2008). The physical  reason for this  change in
slope  is probably  a  combination of  AGN  feedback, and  the  change in  the
efficiency  of radiative  cooling, supernova  feedback and  dynamical friction
(e.g.  Lin  \etal 2004; Cooray  \& Milosavljevi\'c 2005).  For  comparison, we
also  show in the  upper-left panel  the mean  $L_c-M_h$ relation  obtained by
Cacciato et al. (2008) based on  the CLF models.  The excellent agreement with
our  prediction indicates that  although obtained  via different  methods, the
mean $L_c-M_h$ relation are well constrained in both investigations.

For the  stellar mass of the central  galaxies, fitting the model  to the data
shown in  the right panels of  Fig.  \ref{fig:fit} gives  $\log M_0 =10.3061$,
$\log M_1 =11.0404$,  $\alpha=0.3146$, $\beta=4.5427$, $\sigma_0=0.1730$. Note
that $M_{\ast,c}  \propto M_h^{4.9}$ for $\log  M_h \ll 11.0 $,  which is even
steeper than the $L_c$ - $M_h$ relation.

The above  results are  obtained under  the assumption that  the value  of the
dispersion, $\sigma$,  is independent of  halo mass. Although  consistent with
the  data for  $M_h \ga  10^{12} h^{-2}  \Msun$, and  supported  by satellite
kinematics  and semi-analytical models  (More et  al.  2008),  it is  not well
constrained  in low  mass halos.   We therefore  now test  the impact  of this
assumption on  our results.   For this purpose,  we consider two  models where
$\sigma$ is required  to change from the observed  value at $\log (M_h/\msunh)
\sim 12$  either to $0$ or  to $0.8$ at  $\log (M_h/\msunh) \sim 10$  (see the
dotted  and  dashed  curves  in  the middle  panels  of  Fig.   \ref{fig:fit},
respectively).  The best fits of these two test models are shown as the dotted
and dashed  lines in the other panels.   Clearly, our results for  the $L_c$ -
$M_h$ and $M_c$ - $M_h$ relations  are robust with respect to our assumption
that $\sigma$ is independent of halo mass.

\section{Summary}
\label{sec_summary}

In this paper, we have derived the luminosity and stellar mass functions for
different populations of galaxies (central versus satellite; red versus blue;
and galaxies in halos of different masses), and for groups themselves, using a
large galaxy group catalogue constructed from the SDSS Data Release 4
(DR4). Our main results can be summarized as follows:
\begin{enumerate}
\item  For central  galaxies, the  conditional stellar  mass  function (CSMF),
  which  describes the stellar  mass distribution  of galaxies  in halos  of a
  given mass can be well described  by a log-normal distribution, with a width
  $\sigma_{\rm log M_\ast}\sim 0.17$, quite independent of the host halo mass.
  The  median   central  stellar  mass  increases  rapidly   with  halo  mass,
  $M_\ast\propto M_h^{4.9}$, for halos with masses $M_h\ll 10^{11}\msunh$, but
  only   slowly,   $M_\ast\propto   M_h^{0.3}$,   for   halos   with   $M_h\gg
  10^{13}\msunh$.
\item For satellite  galaxies, the conditional stellar mass  function in halos
  of different masses can be described reasonably well by a modified Schechter
  form than breaks away faster than the Schechter function at the massive end.
  The  faint end  slope appears  to  be steeper  for more  massive halos.   On
  average, there are about 3 times as many central galaxies as satellites.
\item  When stellar  mass functions  are measured  separately for  galaxies of
  different colors,  we find that the  central population is  dominated by red
  galaxies  at the massive  end, and  by blue  galaxies at  the low-mass  end. 
  Among the satellite population, there  are in general more red galaxies than
  blue ones.  At the very  low-mass end ($M_\ast \la 10^9 h^{-2}\Msun$), there
  is a marked increase in the  number of red centrals. We speculate that these
  galaxies are located  close to large halos so that  their star formation has
  been affected by their environments.
\item The stellar-mass  function of galaxy groups, which  describes the number
  density of  galaxy groups as a function  of the total stellar  mass of group
  member  galaxies,  is  well  described   by  a  double  power  law,  with  a
  characteristic stellar mass at  $\sim 4\times 10^{10}h^{-2}\Msun$. This form
  is very different  from that of the halo mass  function, indicating that the
  efficiencies  of  star formation  in  halos  of  different masses  are  very
  different.
\item  The stellar  mass function  for  the central  galaxies can  be used  to
  provide stringent constraint  on the mean $M_{\ast,c}$ -  $M_h$ relation for
  low-mass halos.
\end{enumerate}

We anticipate that a comparison of these results with predictions of numerical
simulations and/or  semi-analytical models will  provide stringent constraints
on how galaxies form and evolve in dark matter halos.


\section*{Acknowledgments}

We are  grateful to the anonymous  referee for useful  and insightful comments
that greatly helped to improve the presentation of this paper. XY acknowledges
the  University   of  Massachusetts  for  hospitality  where   this  work  was
finalized. This  work is supported by  the {\it One  Hundred Talents} project,
Shanghai Pujiang  Program (No.   07pj14102), 973 Program  (No.  2007CB815402),
the CAS Knowledge Innovation Program (Grant No.  KJCX2-YW-T05) and grants from
NSFC (Nos.  10533030,  10673023, 10821302). HJM would like  to acknowledge the
support of NSF AST-0607535, NASA AISR-126270 and NSF IIS-0611948.


\label{lastpage}

\end{document}